\documentclass[10pt]{book}
\usepackage{natbib}
\usepackage{amsmath,graphicx,amssymb,verbatim}
\bibpunct{(}{)}{;}{a}{}{,}
\usepackage{array}
\newcommand{\example}[1]{\textbf{Example #1. }}
\newcommand{\qed}{{\unskip\nobreak\hfil\penalty50\hskip2em\vadjust{}
            \nobreak\hfil$\Box$\parfillskip=0pt\finalhyphendemerits=0\par}}

\newtheorem{thm}{Theorem}[chapter] 
\newtheorem{lemma}{Lemma}[chapter] 
\newtheorem{cor}{Corollary}[chapter]

\begin{document}
\bibliographystyle{sinica}

\renewcommand{\baselinestretch}{1.2}
\markright{
}
\markboth{\hfill{\footnotesize\rm X. JESSIE JENG AND Z. JOHN DAYE 
}\hfill}
{\hfill {\footnotesize\rm COVARIANCE THRESHOLDING FOR VARIABLE SELECTION 
} \hfill}
\renewcommand{\thefootnote}{}
$\ $\par \fontsize{10.95}{14pt plus.8pt minus .6pt}\selectfont
\vspace{0.8pc} \centerline{\large\bf SPARSE COVARIANCE
THRESHOLDING FOR
} \vspace{2pt} \centerline{\large\bf HIGH-DIMENSIONAL VARIABLE
SELECTION
} \vspace{.4cm} \centerline{X. Jessie Jeng and Z. John Daye} \vspace{.4cm} \centerline{\it Purdue University}
\vspace{.55cm} \fontsize{9}{11.5pt plus.8pt minus .6pt}\selectfont

\begin{quotation}
\noindent {\it Abstract:} In high-dimensions, many variable
selection methods, such as the lasso, are often limited by
excessive variability and rank deficiency of the sample covariance
matrix.  Covariance sparsity is a natural phenomenon in
high-dimensional applications, such as microarray analysis, image
processing, etc., in which a large number of predictors are
independent or weakly correlated.  In this paper, we propose the
covariance-thresholded lasso, a new class of regression methods
that can utilize covariance sparsity to improve variable
selection.  We establish theoretical results, under the random
design setting, that relate covariance sparsity to variable
selection. Real-data and simulation examples indicate that our
method can be useful in improving variable selection
performances.\par

\vspace{9pt} \noindent {\it Key words and phrases:} Consistency,
covariance sparsity, large p small n, random design, regression,
regularization.
\par
\end{quotation}\par

\fontsize{10.95}{14pt plus.8pt minus .6pt}\selectfont
\setcounter{chapter}{1}
\noindent {\bf 1. Introduction}
\par \smallskip
Variable selection in high-dimensional regression is a central
problem in Statistics and has stimulated much interest in the past
few years.
Motivation for developing effective variable selection methods in
high-dimensions comes from a variety of applications, such as gene
microarray analysis, image processing, etc., where it is necessary
to identify a parsimonious subset of predictors to improve
interpretability and prediction accuracy.
In this paper, we consider the following linear model for $X =
(X_1, X_2, \ldots, X_{p})^T$ a vector of $p$ predictors and $Y$ a
response variable,
\begin{equation}\label{LinMod}
Y = X \beta^{*} + \epsilon,
\end{equation}
where $\beta^* = (\beta_1^*, \beta_2^*, \ldots, \beta_{p}^*)^T$ is
a vector of regression coefficients and $\epsilon$ is a normal
random error with mean 0 and variance $\sigma^2$. If $\beta_j^*$
is nonzero, then $X_j$ is said to be a true variable; otherwise,
it is an irrelevant variable. Further, when only a few
coefficients $\beta^*_j$'s are believed to be nonzero, we refer to
(\ref{LinMod}) as a sparse linear model. The purpose of variable
selection is to separate the true variables from the irrelevant
ones based upon some observations of the model. In many
applications, $p$ can be fairly large or even larger than $n$. The
problem of large $p$ and small $n$ presents a
fundamental challenge for variable selection.  

Recently, various methods based upon $L_1$ penalized least squares
are proposed for variable selection. The lasso, introduced by
\citet*{Tibshirani96}, is the forerunner and foundation for many
of these methods. Suppose that $\textbf{y}$ is an $n\times 1$
vector of observed responses centered to have mean 0 and
$\mathbf{X}=(\mathbf{X}_1,\mathbf{X}_2,\ldots,\mathbf{X}_p)$ is an
$n\times p$ data matrix with each column $\mathbf{X}_j$
standardized to have mean zero and variance of 1. We may
reformulate the lasso as the following,
\begin{equation}\label{LasEqu}
\hat \beta^{Lasso}(\lambda_n) =  \arg\min_{\beta} \left\{ \beta^T
\hat{\mathbf{\Sigma}} \beta - 2 \beta^T \left({1\over n}
\mathbf{X}^T \mathbf{y} \right)+ 2 \lambda_n
\left\|\beta\right\|_1 \right\},
\end{equation}
where $\hat{\mathbf{\Sigma}}=\mathbf{X}^T \mathbf{X}/n$ is the
sample covariance or correlation matrix. Consistency in variable
selection for the lasso has been proved under the neighborhood
stability condition in \citet*{Meinshausen06} and under the
irrepresentable condition in \citet*{Zhao06}.
Compared with traditional variable selection procedures, such as
all subset selection, AIC, BIC, etc., the lasso has continuous
solution paths and can be computed efficiently using innovative
algorithms, such as the LARS in \citet*{Efron04}.
Since its introduction, the lasso has emerged as one of the most
widely-used methods for variable selection.

In the lasso literature, data matrix $\mathbf{X}$ is often assumed
to be fixed. However, this assumption may not be realistic in
high-dimensional applications, where data usually come from
observational rather than experimental studies. In this paper, we
assume the predictors $X_1,X_2,\ldots,X_p$ in (\ref{LinMod}) to be
random with $E(X)=0$ and $E(X X^T) =
\mathbf{\Sigma}=(\sigma_{ij})_{1\le i \le p, 1\le j \le p}$.
In addition, we assume that the population covariance matrix
$\mathbf{\Sigma}$ is sparse in the sense that the proportion of
nonzero $\sigma_{ij}$ in $\mathbf{\Sigma}$ is relatively small.
Motivations for studying sparse covariance matrices come from a
myriad of applications in high-dimensions, where a large number of
predictors can be independent or weakly correlated with each
other.
For example, in gene microarray analysis, it is often reasonable
to assume that genes belonging to different pathways or systems
are independent or weakly correlated \citep*{Rothman08,Wagaman08}.
In these applications, the number of nonzero covariances in
$\mathbf{\Sigma}$ can be much smaller than $p(p-1)/2$, the total
number of covariances.

An important component of lasso regression (\ref{LasEqu}) is the
sample covariance matrix $\hat{\mathbf{\Sigma}}$. We note that the
sample covariance matrix is rank-deficient when $p>n$. This can
cause the lasso to saturate after at most $n$ variables are
selected. Moreover, the `large $p$ and small $n$' scenario can
cause excessive variability of sample covariances between the true
and irrelevant variables. 
This deteriorates the ability of the lasso to separate true
variables from irrelevant ones. More specifically, a sufficient
and almost necessary condition for the lasso to be variable
selection consistent is derived in \citet*{Zhao06}, which they
call the irrepresentable condition.  It poses constraint on the
inter-connectivity between the true and irrelevant variables in
the following way. Let $S = \{j \in \{1, \ldots, p \} \mid
\beta^*_j \ne 0 \}$ and $C=\{1, 2, \ldots, p\}-S$, such that $S$
is the collection of true variables and $C$ is the complement of
$S$ that is composed of the irrelevant variables. Assume that the
cardinality of $S$ is $s$; in other words, there are $s$ true
variables and $p-s$ irrelevant ones. Further, let $\mathbf{X}_S$
and $\mathbf{X}_C$ be sub-data matrices of $\mathbf{X}$ that
contain the observations of the true and irrelevant variables,
respectively. Define $\hat{\mathcal{I}}=\mid{
\hat{\mathbf{\Sigma}}_{CS}
(\hat{\mathbf{\Sigma}}_{SS})^{-1}sgn(\beta^*_S)}\mid$, where
$\hat{\mathbf{\Sigma}}_{CS}=\mathbf{X}_{C}^T \mathbf{X}_{S}/n$ and
$\hat{\mathbf{\Sigma}}_{SS}=\mathbf{X}_S^T \mathbf{X}_S/n$. We
refer to $\hat{\mathcal{I}}$ as the {\em sample irrepresentable
index}. It can be interpreted as representing the amount of
inter-connectivity between the true and irrelevant variables.
In order for lasso to select the true variables consistently,
irrepresentable condition requires $\hat{\mathcal{I}}$ to be
bounded from above, that is $\hat{\mathcal{I}}<1-\epsilon$ for
some $\epsilon \in (0,1)$, entry-wise. Clearly, excessive
variability of the sample covariance matrix induced by large $p$
and small $n$ can cause $\hat{\mathcal{I}}$ to exhibit large
variation that makes the irrepresentable condition less likely to
hold. These inadequacies motivate us to consider alternatives to
the sample covariance matrix to improve variable selection for the
lasso in high-dimensions.

Next, we provide some insight on how the sparsity of the
population covariance matrix can influence variable selection for
the lasso. Under random design assumption on $X$, the
inter-connectivity between the true and irrelevant variables can
be stated in terms of their population variances and covariances.
Let $\mathbf{\Sigma}_{CS}$ be the covariance matrix between the
irrelevant variables and true variables and $\mathbf{\Sigma}_{SS}$
the variance-covariance matrix of the true variables. We define
the {\em population irrepresentable index} as $ \mathcal{I}=\mid
\mathbf{\Sigma}_{CS}\mathbf{\Sigma}_{SS}^{-1}sgn{(\beta^*_S)}\mid.$
Intuitively, 
the sparser the population covariances $\mathbf{\Sigma}_{CS}$ and
$\mathbf{\Sigma}_{SS}$ are, or the sparser $\mathbf{\Sigma}$ is,
the more likely that $\mathcal{I} < 1-\epsilon$, entry-wise.
This property, however, does not automatically trickle down to the
sample irrepresentable index $\hat{\mathcal{I}}$, due to its
excessive variability. When $\mathbf{\Sigma}_{CS}$ and
$\mathbf{\Sigma}_{SS}$ are known a priori to be sparse and
$\mathcal{I}<1-\epsilon$, entry-wise, some regularization on the
covariance can be used to reduce the variabilities of
$\hat{\mathbf{\Sigma}}$ and $\hat{\mathcal{I}}$ and allow the
irrepresentable condition to hold more easily for
$\hat{\mathcal{I}}$.
Furthermore, the sample covariance matrix $\hat{\mathbf{\Sigma}}
=\mathbf{X}^T \mathbf{X}/n$ is obviously non-sparse; and imposing
sparsity on $\hat{\mathbf{\Sigma}}$ has the benefit of sometimes increasing the rank of the sample covariance matrix.

We use an example to demonstrate how rank deficiency and excessive
variability of the sample covariance matrix
$\hat{\mathbf{\Sigma}}$ can compromise the performance of the
lasso for large $p$ and small $n$. Suppose there are 40 variables
($p=40$) and $\mathbf{\Sigma}=I_p$ ( $I_p$ is the $p\times p$
identity matrix).  Since all variables are independent of each
other, the population irrepresentable index clearly satisfies
$\mathcal{I}<1-\epsilon$, entry-wise. Further, we let
$\beta^*_j=2$, for $1\le j \le 10$, and $\beta^*_j=0$, for $11\le
j \le 40$. The error standard deviation $\sigma$ is set to be
about 6.3 to have a signal-to-noise ratio of approximately 1. The
lasso, in general, does not take into consideration the structural
properties of the model, such as the sparsity or the orthogonality
of $\mathbf{\Sigma}$ in this example. One way to take advantage of
the orthogonality of $\mathbf{\Sigma}$ is to replace
$\hat{\mathbf{\Sigma}}$ in (\ref{LasEqu}) by $I_p$, which leads to
the univariate soft thresholding (UST) estimates
$\hat{\beta}^{UST}_j=sgn(r_j)(\mid r_j-\lambda\mid)^{+}$, where
$r_j=\mathbf{X}_j^T Y/n$ for $1\le j \le p$. We compare the
performances of the lasso and UST over various sample sizes ($5\le
n \le 250$) using the variable selection measure $G$. $G$ is defined as the geometric mean between sensitivity, $\textrm{(no. of true positives)}/s$, and specificity, $1-(\textrm{no. of false positives})/(p-s)$ \citep*{Tibshirani05,Chong05,Kubat98}. $G$ varies between 0 and 1. Larger $G$ indicates better selection with a
larger proportion of variables classified correctly.

\begin{figure}[!h]\renewcommand{\baselinestretch}{1}
\centering
\includegraphics[width = 2.5 in, angle=0]{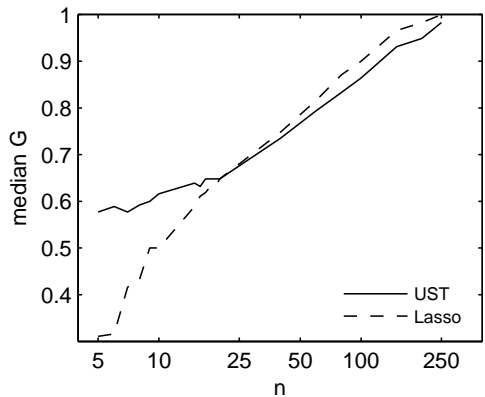}
\caption{Median G (from $n$=5 to $n$=250) for illustrating example
based upon 200 replications.} \label{fig:intro}
\end{figure}

Figure \ref{fig:intro} plots the median $G$ based on 200
replications for the lasso and UST against sample sizes. For each
replication, $\lambda$ is determined ex post facto by the optimal
$G$ in order to avoid stochastic errors from tuning parameter
estimation, such as by using cross-validation.
It is clear from Figure \ref{fig:intro} that, when $n>20$, lasso
slightly outperforms UST; when $n<20$, the performance of lasso
starts to deteriorate precipitously, whereas the performance of
UST declines at a much slower pace and starts to outperform lasso.
This example suggests that when $p$ is large and $n$ is relatively
small, sparsity of $\mathbf{\Sigma}$ can be used to enhance
variable selection.

The discussions
above motivate us to consider improving the performance of the
lasso by applying regularization to the sample covariance matrix
$\hat{\mathbf{\Sigma}}$. A good sparse covariance-regularizing
operator on $\hat{\mathbf{\Sigma}}$ should satisfy the following
properties:
\begin{enumerate}\setlength{\itemsep}{-.15\baselineskip}
\item The operator \emph{stabilizes} $\hat{\mathbf{\Sigma}}$.

\item The operator can \emph{increase the rank} of
$\hat{\mathbf{\Sigma}}$.

\item The operator \emph{utilizes the underlying sparsity} of the
covariance matrix.
\end{enumerate}
The first and second properties are obviously useful and have been
explored in the literature. For example,
the elastic net, introduced in \citet*{Zou05}, replaces
$\hat{\mathbf{\Sigma}}$ by $\hat{\mathbf{\Sigma}}_{EN} =
(\hat{\mathbf{\Sigma}}+\lambda_2 I)/(1+\lambda_2)$ in
(\ref{LasEqu}), where $\lambda_2>0$ is a tuning parameter.
$\hat{\mathbf{\Sigma}}_{EN}$ can be more stable and have higher
rank than $\hat{\mathbf{\Sigma}}$ but is non-sparse.
Nonetheless, in many applications, utilizing the underlying
sparsity may be more crucial in improving the lasso when data is
scarce, such as under the large $p$ and small $n$ scenario.

Recently, various regularization methods have been proposed in the
literature for estimating high-dimensional variance-covariance
matrices. Some examples include tapering proposed by
\citet*{Furrer07}, banding by \citet*{BL08}, thresholding by
\citet*{Bickel08} and \citet*{El07}, and generalized thresholding
by \citet*{Rothman08}. We note that covariance thresholding
operators can satisfy all three properties outlined in the
previous paragraph; in particular, they can generate sparse
covariance estimates to accommodate for the covariance sparsity
assumption.
In this paper, we propose to apply covariance-thresholding on the
sample covariance matrix $\hat{\mathbf{\Sigma}}$ in (\ref{LasEqu})
to stabilize and improve the performances of the lasso. We call
this procedure the \emph{covariance-thresholded lasso}. We
establish theoretical results that relate the sparsity of the
covariance matrix with variable selection
and compare them to those of the lasso. Simulation and real-data
examples are reported. Our results suggest that
covariance-thresholded lasso can improve upon the lasso, adaptive
lasso, and elastic net, especially when $\mathbf{\Sigma}$ is
sparse, $n$ is small, and $p$ is large. Even when the underlying
covariance is non-sparse, covariance-thresholded lasso is still
useful in providing robust variable selection in high-dimensions.

\citet*{Witten08} has recently proposed the scout
procedure, that applies regularization to the inverse covariance or precision matrix. We note that this is quite different from the covariance-thresholded lasso that regularizes the sample
covariance matrix $\hat{\mathbf{\Sigma}}$ directly.  Furthermore, the scout penalizes using the matrix norm $\|{\mathbf{\Theta}}_{\mathbf{X}\mathbf{X}}\|^{p}_{p}=\sum_{ij} |{\theta}_{ij}|^p $, where $\mathbf{\Theta}$ is an estimate of $\mathbf{\Sigma}^{-1}$, whereas the covariance-thresholded lasso regularizes individual covariances $\hat{\sigma}_{ij}$ directly.  In our results, we will show that the scout is potentially very similar to the elastic net and that the covariance-thresholded lasso can often outperform the scout in terms of variable selection for $p > n$.

The rest of the paper is organized as follows. In Section 2, we
present covariance-thresholded lasso in detail and a modified LARS
algorithm for our method. We discuss a generalized class of
covariance-thresholding operators and explain how
covariance-thresholding can stabilize the LARS algorithm for the
lasso. In Section 3, we establish theoretical results on variable
selection for the covariance-thresholded lasso. The effect of
covariance sparsity on variable selection is especially
highlighted.
In Section 4, we provide simulation results of
covariance-thresholded lasso at $p > n$, and, in Section 5, we
compare the performances of covariance-thresholded lasso with
those of the lasso, adaptive lasso, and elastic net using 3
real-data sets. Section 6 concludes with further discussions and
implications.

\par \bigskip

\setcounter{chapter}{2}
\noindent {\bf 2. The Covariance-Thresholded Lasso}
\par \smallskip
Suppose that the response $\mathbf{y}$ is centered and each column
of the data matrix $\mathbf{X}$ is standardized, as in the lasso
(\ref{LasEqu}). We define the covariance-thresholded
lasso 
estimate as
\begin{equation} \label{def:CT-Lasso}
\hat \beta^{CT-Lasso}(\nu, \lambda_n) =  \arg\min_{\beta} \left\{
\beta^T \hat{\mathbf{\Sigma}}_\nu \beta - 2 \beta^T \left({1\over
n}\mathbf{X}^T \mathbf{y} \right)+ 2 \lambda_n
\left\|\beta\right\|_1 \right\},
\end{equation}
where $\hat{\mathbf{\Sigma}}_\nu \equiv [\hat \sigma_{ij}^\nu]$,
$\hat \sigma_{ij}^\nu = s_\nu(\hat \sigma_{ij})$, $\hat
\sigma_{ij} = \sum_{i = 1}^n X_{ki} X_{kj}/n$, and $s_\nu(\cdot)$
is a pre-defined covariance-thresholding
operator with $0 \le \nu < 1$. 
If the identity function is used as the covariance-thresholding
operator, that is $s_\nu(x)=x$ for any $x$, then $\hat
\beta^{CT-Lasso}(\nu, \lambda_n) =\hat \beta^{Lasso}$.
\par \bigskip

\noindent {\bf 2.1. Sparse Covariance-thresholding Operators}
\par \smallskip
We consider a generalized class of covariance-thresholding
operators $s_\nu(\cdot)$ introduced in \citet*{Rothman08}.  These
operators satisfy the following properties,
\begin{equation} \label{property:s}
s_\nu(\hat{\sigma}_{ij}) = 0  \text{~for~} |\hat{\sigma}_{ij}| \le
\nu, \quad |s_\nu(\hat{\sigma}_{ij})| \le |\hat{\sigma}_{ij}|,
\quad |s_\nu(\hat{\sigma}_{ij}) - \hat{\sigma}_{ij}| \le \nu.
\end{equation}
The first property enforces sparsity for covariance estimation;
the second allows shrinkage of covariances; and the third limits
the amount of shrinkage. These operators satisfy the desired
properties outlined in the Introduction 
for sparse covariance-regularizing operators and
represent a wide spectrum of thresholding procedures that can
induce sparsity and stabilize the sample covariance matrix. In
this paper, we will consider the following covariance-thresholding
operators
for $\hat{\sigma}_{ij}$ when $i\neq j$.  
\begin{align}
  & \text{1. Hard thresholding:} \quad s_\nu^{\textnormal{Hard}}(\hat{\sigma}_{ij}) = \hat{\sigma}_{ij}
1(|\hat{\sigma}_{ij}| > \nu). \label{eqn:hard}\\
  & \text{2. Soft thresholding:} \quad s_\nu^{\textnormal{Soft}}(\hat{\sigma}_{ij}) =
sgn(\hat{\sigma}_{ij}) (|\hat{\sigma}_{ij}| - \nu)^+. \label{eqn:soft}\\
  & \text{3. Adaptive thresholding:} \quad \text{For }\gamma \ge 0,
  \nonumber\\
  & \qquad \qquad s_\nu^{\textnormal{Adapt}}(\hat{\sigma}_{ij}) = sgn(\hat{\sigma}_{ij}) (|\hat{\sigma}_{ij}| - \nu^{\gamma+1}|\hat{\sigma}_{ij}|^{-\gamma})^+. \label{eqn:adaptive}
\end{align}
The above operators are used in \citet*{Rothman08} for estimating
variance-covariance matrices, and
it is easy to check that they satisfy the properties in
(\ref{property:s}).

\begin{figure}[!tb]\renewcommand{\baselinestretch}{1}
  \centering
  \begin{tabular}{cccc}
  CT Hard & CT Soft & CT Adapt & Elastic Net \\
  \includegraphics[width=.2\textwidth]{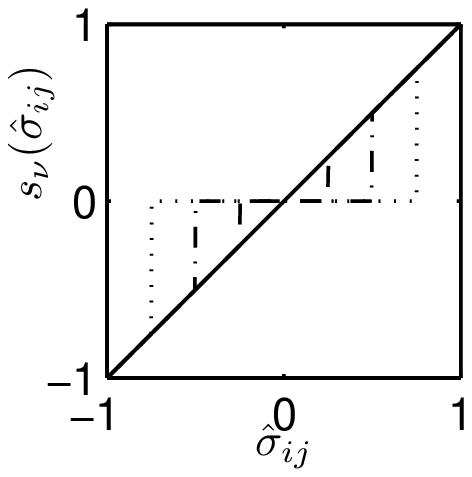}&
  \includegraphics[width=.2\textwidth]{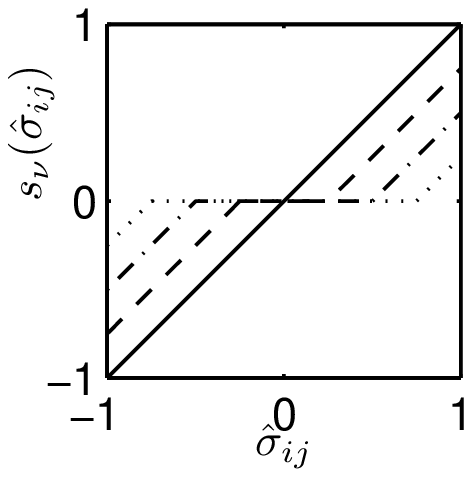}&
  \includegraphics[width=.2\textwidth]{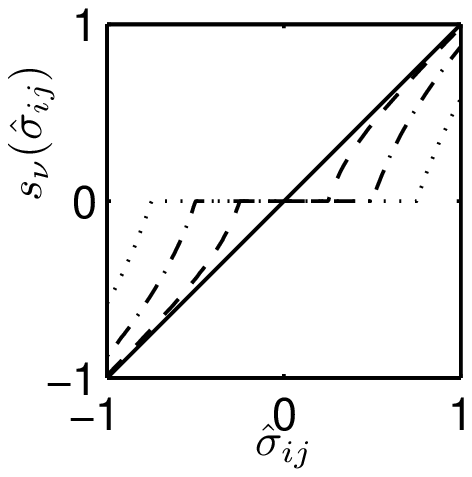}&
  \includegraphics[width=.2\textwidth]{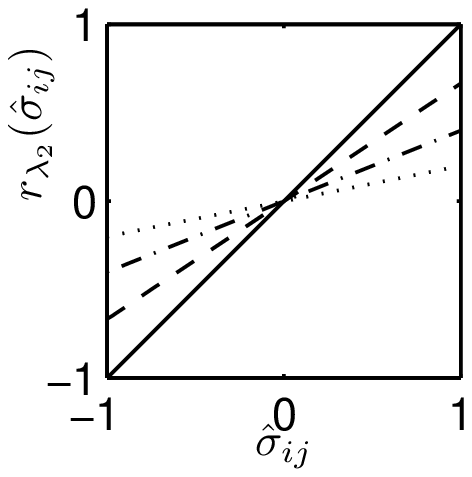}\\
  \end{tabular}
  \caption{Hard, soft, adaptive ($\gamma$=2)
sparse covariance-thresholding
  operators with $\nu$ varying over 0 (solid), 0.25 (dashed), 0.5 (dot-dashed),
  and 0.75 (dotted); and elastic net covariance-regularizing operator with $\lambda_2$ varying over 0 (solid), 0.5 (dashed), 1.5 (dot-dashed), and 4 (dotted).
  }
  \label{fig:operators}
\end{figure}

In Figure \ref{fig:operators}, we depict the sparse
covariance-thresholding operators
(\ref{eqn:hard}-\ref{eqn:adaptive}) for varying $\nu$.  Hard
thresholding presents a discontinuous thresholding of covariances,
whereas soft thresholding offers continuous shrinkage.
Adaptive thresholding presents less regularization on covariances
with large magnitudes than soft thresholding.

Figure \ref{fig:operators} further includes the elastic net
covariance-regularizing operator, $r_{\lambda_2}(\hat{\sigma}_{ij})=(\hat{\sigma}_{ij} +
\lambda_2)/(1+\lambda_2)$ for $i \neq j$.  Apparently, this operator is non-sparse and does not satisfy the first property in (\ref{property:s}).  In particular, we see that the elastic net penalizes covariances with large magnitudes more severely than those with small magnitudes.  In some situations, this has the benefit of
alleviating multicollinearity as it shrinks covariances of highly
correlated variables.  However, under high-dimensionality and when much of the random perturbation of the covariance matrix arises from small but numerous covariances, the elastic net in attempting to control these variabilities may inadvertently penalize covariances with large magnitudes severely, which may introduce large bias in estimation and compromise the performance of the elastic net under some scenarios.

\par \bigskip

\noindent {\bf 2.2. Computations}
\par \smallskip
The lasso solution paths are shown to be piecewise linear in
\citet*{Efron04} and \citet*{Rosset07}.  This property allows
\citet*{Efron04} to propose the efficient LARS algorithm for the
lasso. Likewise, in this section, we propose a piecewise-linear algorithm for the covariance-thresholded lasso.

\begin{figure}[!tb]\renewcommand{\baselinestretch}{1}
  \centering
  \begin{tabular}{cc}
  (a) Example 1 & (b) Example 2\\
  \includegraphics[width=3.in]{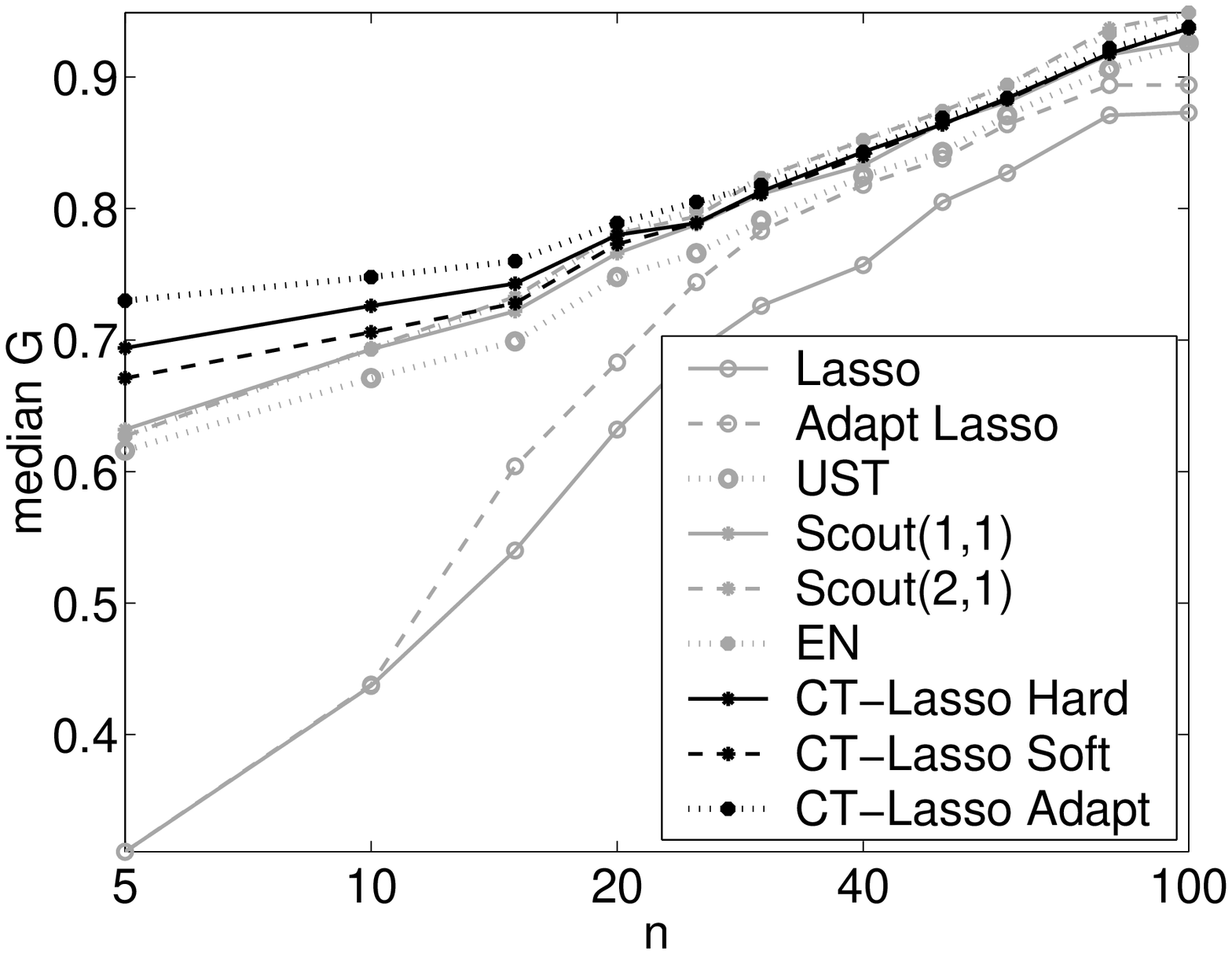}&
  \includegraphics[width=3.in]{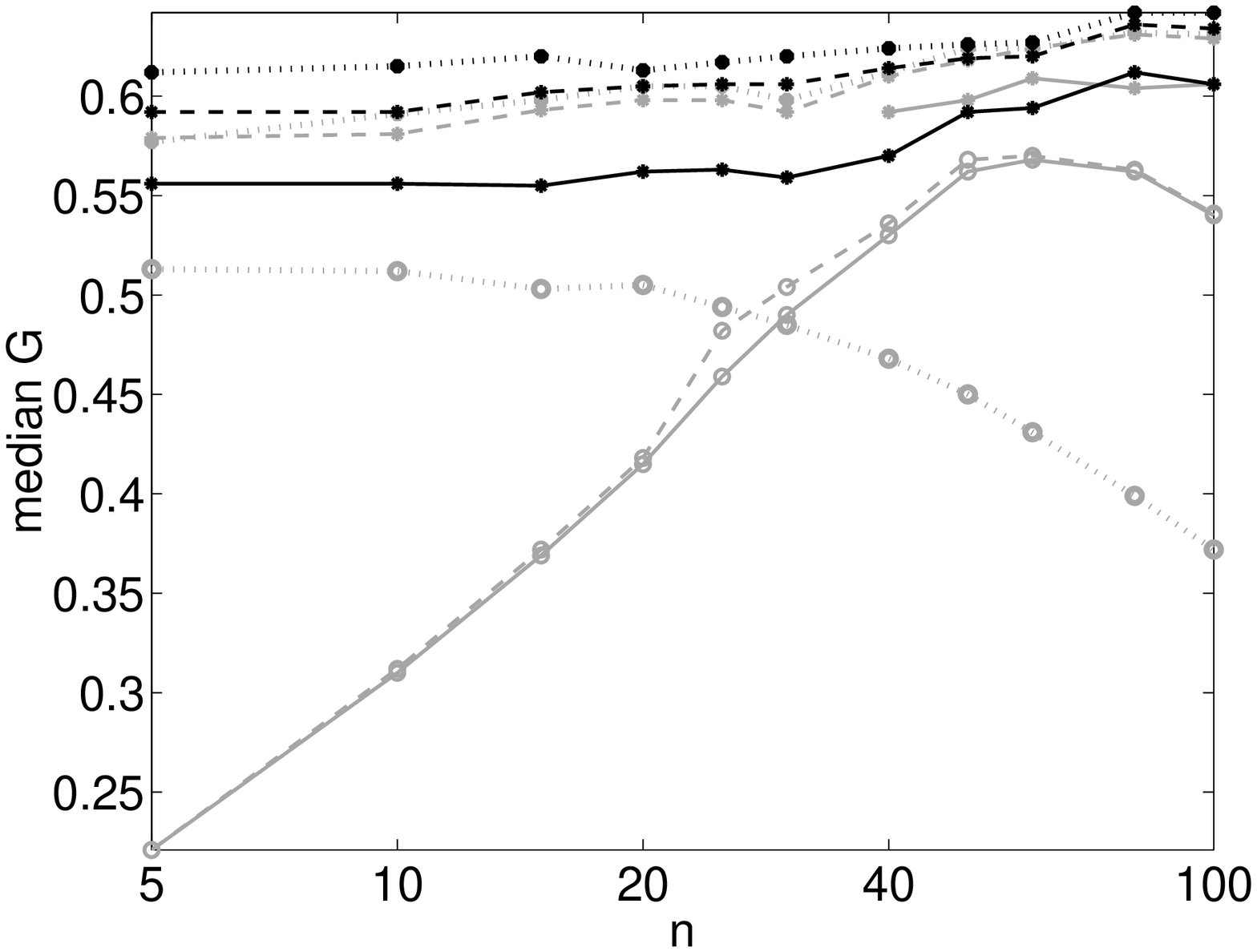}\\
  \multicolumn{2}{c}{(c) Example 3}\\
  \multicolumn{2}{c}{
  \includegraphics[width=3.in]{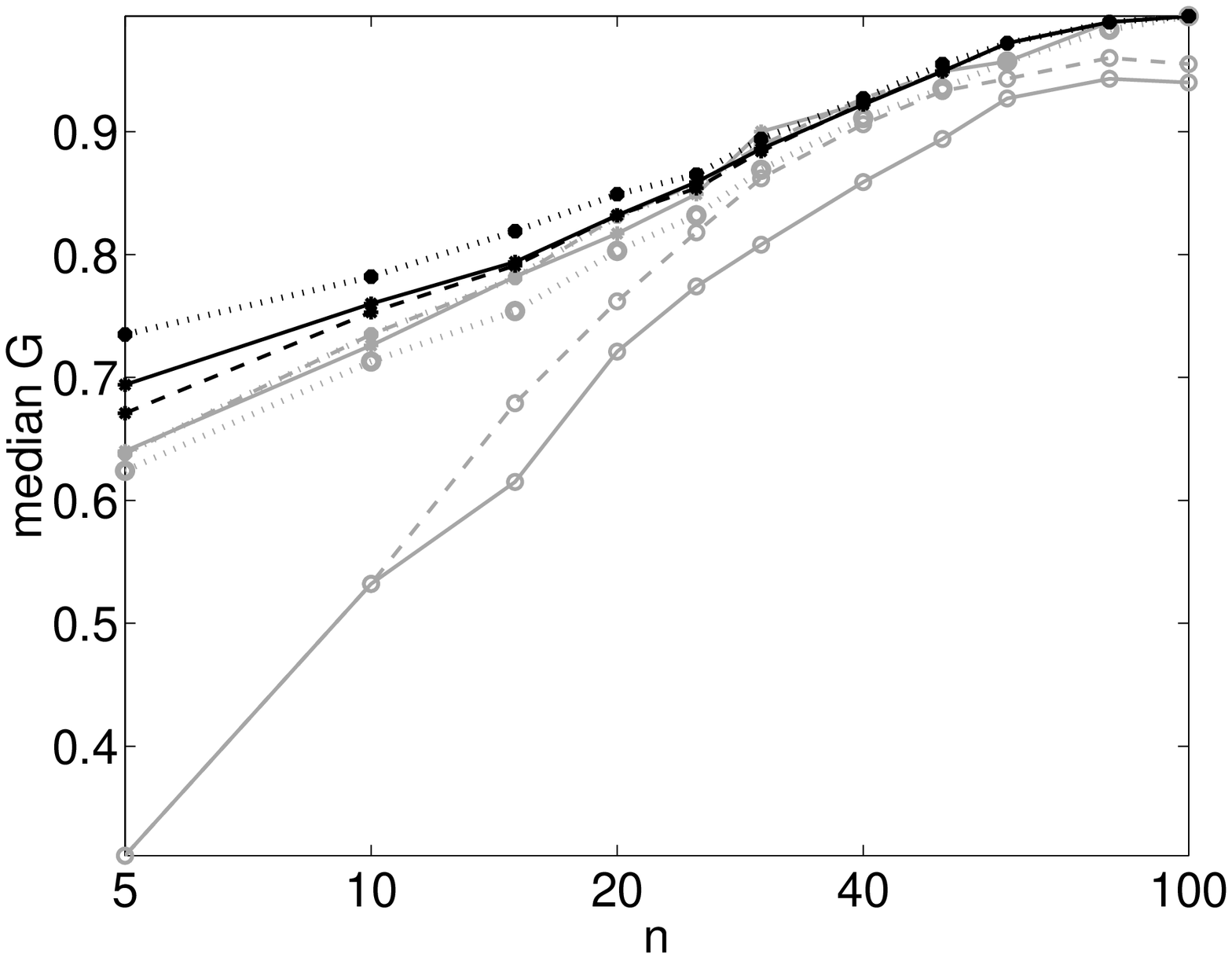}}\\
  \end{tabular}
  \caption{Variable selection performances using best-possible selection
  of tuning parameters based on 200 replications
  at $n=\{5, 10, 15, 20, 25, 30, 40,
  50, 60, 80, 100\}$.
  }\label{fig:best}
\end{figure}

We note that the loss function $\beta^T \hat{\mathbf{\Sigma}}_\nu \beta - 2 \beta^T \mathbf{X}^T \mathbf{y}/n$ in (\ref{def:CT-Lasso}) can sometimes be non-convex since $\hat{\mathbf{\Sigma}}_\nu$ may possess negative eigenvalues for some $\nu$.  This usually may occur for intermediary values of $\nu$, as $\hat{\mathbf{\Sigma}}_\nu$ is at least semi-positive definite for $\nu$ close to 0 or 1.  Furthermore, we note that the penalty $2 \lambda_n \|\beta\|_1$ is a convex function and dominates in (\ref{def:CT-Lasso}) for $\lambda_n$ large.  Intuitively, this means that the optimization problem for covariance-thresholded lasso is almost convex for $\beta$ sparse.  This is stated conservatively in the following theorem by using second-order condition from nonlinear programming \citep{McCormick76}.
\begin{thm}\label{thm:piecewiseLinear}
Let $\nu$ be fixed.  If $\hat{\mathbf{\Sigma}}_\nu$ is semi-positive definite, the covariance-thresholded lasso solutions
$\hat{\beta}^{CT-Lasso}(\nu, \lambda_n)$ for (\ref{def:CT-Lasso})
are piecewise linear with respect to $\lambda_n$.  If $\hat{\mathbf{\Sigma}}_\nu$ possesses negative eigenvalues, a set of covariance-thresholded lasso solutions, which may be local minima for (\ref{def:CT-Lasso}) under strict complementarity, is piecewise linear with respect to $\lambda_n$ for $\lambda_n \ge \lambda^*$,
where $\lambda^*= \min \{\lambda>0 : \text{ sub-matrix } (\hat{\mathbf{\Sigma}}_\nu)_\mathcal{A} \text{ remains positive definite for } \mathcal{A} = \{j: \hat
\beta_j^{CT-Lasso} (\nu, \lambda) \ne 0 \} \}$
\end{thm}
The proof for Theorem \ref{thm:piecewiseLinear} is outlined in Appendix 7.6.  Strict complementarity, described in Appendix 7.6, is a technical condition that allows the second-order condition to be more easily interpreted and usually holds with high probability. We note that, when $\hat{\mathbf{\Sigma}}_\nu$ has negative eigenvalues, the solution $\hat{\beta}^{CT-Lasso}(\nu,\lambda_n)$ is global if $|\mathbf{x}_j^T \mathbf{y}/n| < \lambda_n $ for all $j \notin \mathcal{A}_n= \{j : \hat{\beta}_j^{CT-Lasso}(\nu, \lambda_n) \neq 0\}$ and $(\hat{\mathbf{\Sigma}}_\nu)_{\mathcal{A}_n}$ is positive definite.  Theorem \ref{thm:piecewiseLinear} suggests that piecewise linearity of the covariance-thresholded lasso solution path sometimes may not hold for some $\nu$ when $\lambda_n$ is small, even if a solution may well exist.  This restricts the sets of tuning parameters ($\nu$, $\lambda_n$) for which we can compute the solutions of covariance-thresholded lasso efficiently using a LARS-type algorithm.  We note that the elastic net does not suffer from a potentially non-convex optimization.  However, as we will demonstrate in Figure \ref{fig:best} of Section 4, covariance-thresholded lasso with restricted sets of ($\nu$, $\lambda_n$) is, nevertheless, rich enough to dominate the elastic net in many situations.

Theorem \ref{thm:piecewiseLinear} establishes that a set of covariance-thresholded lasso solutions are piecewise linear.  This further provides us with an efficient modified LARS algorithm for computing the covariance-thresholded lasso.
Let
\begin{equation} \label{eqn:LARSc}
(\hat{c}_\nu)_j = \frac{1}{n} \mathbf{X}_j^T \mathbf{y} -
(\hat{\mathbf{\Sigma}}_\nu)^T_j \beta
\end{equation}
be estimates for the covariate-residual correlations $c_j$. Further, we denote the minimum eigenvalue of $A$ as $\Lambda_{min}(A)$. The covariance-thresholded lasso can be computed with the following algorithm.

\begin{enumerate}\setlength{\itemsep}{-.15\baselineskip} 
    \item[] ALGORITHM: Covariance-thresholded LARS
    \emph{\item Initialize $\hat{\mathbf{\Sigma}}_\nu$ such that $\hat{\sigma}_{ij}^\nu = s_\nu(\hat{\sigma}_{ij})$,
    $\beta=0$, and  $\hat{\mathbf{c}}_\nu = \frac{1}{n}\mathbf{X}^T \mathbf{y}$.  Let $\mathcal{A} = \arg \max_j |(\hat{c}_\nu)_j|$,
    $\hat{C} = \max |(\hat{\mathbf{c}}_\nu)_\mathcal{A}|$,
    $\mathbf{\gamma}_{\mathcal{A}} = sgn((\hat{\mathbf{c}}_\nu)_{\mathcal{A}})$,
    $\mathbf{\gamma}_{\mathcal{A}^C} = 0$, and $\mathbf{a} = (\hat{\mathbf{\Sigma}}_\nu)^T \gamma$.
    \item Let $\delta_1 = \min^+_{j \in \mathcal{A}} \{-\frac{\beta_j}{\gamma_j}\}$ and $\delta_2 = \min^+_{j \in
    \mathcal{A}^C} \{\frac{\hat{C}-(\hat{c}_\nu)_j}{a_i-a_j},
    \frac{\hat{C}+(\hat{c}_\nu)_j}{a_i+a_j}\}$ for any $i \in
    \mathcal{A}$, where $\min^+$ is taken only over positive
    elements.
    \item Let $\delta = \min(\delta_1, \delta_2)$, $\beta \leftarrow \beta +
    \delta\mathbf{\gamma}$, $\hat{\mathbf{c}}_\nu \leftarrow \hat{\mathbf{c}}_\nu - \delta \textbf{a}$, and
    $\hat{C} = \max_{j \in \mathcal{A}} |(\hat{c}_\nu)_j|$.
    \item If $\delta=\delta_1$, remove the variable hitting $0$
    at $\delta$ from $\mathcal{A}$. If $\delta=\delta_2$, add the variable first attaining equality at $\delta$ to
    $\mathcal{A}$.
    \item Compute the new direction, $\mathbf{\gamma}_\mathcal{A} =
    (\hat{\mathbf{\Sigma}}_\nu)^{-1}_\mathcal{A} sgn(\beta_{\mathcal{A}})$ and $\mathbf{\gamma}_{\mathcal{A}^C}=0$, and let $\mathbf{a} = (\hat{\mathbf{\Sigma}}_\nu)^T \gamma$.
    \item Repeat steps 2-5 until $\min_{j\in\mathcal{A}} |(\hat{c}_\nu)_j| <
    0$ or $\Lambda_{min}({(\hat{\mathbf{\Sigma}}_\nu)_\mathcal{A}})\le 0$.}
\end{enumerate}

The covariate-residual correlations $c_j$ are the most crucial for
computing the solution paths. It determines the variable to be
included at each step and relates directly to the tuning parameter
$\lambda_n$. In the original LARS for the lasso, $c_j$ is
estimated as $\mathbf{X}_j^T \mathbf{y}/n -
\hat{\mathbf{\Sigma}}_j^T\beta$, which uses the sample covariance
matrix $\hat{\mathbf{\Sigma}}$ without thresholding. In
covariance-thresholded LARS, $(\hat{c}_\nu)_j$ is defined using
the covariance-thresholded
estimate $(\hat{\mathbf{\Sigma}}_{\nu})^T_j=
(\hat{\sigma}_{1j}^\nu, \hat{\sigma}_{2j}^\nu, \ldots,
\hat{\sigma}_{pj}^\nu)$, which may contain many zeros. We note
that, in (\ref{eqn:LARSc}), zero-valued covariances
$\hat{\sigma}^{\nu}_{ij}$ have the effect of essentially removing
the associated coefficients from $\beta$, providing parsimonious
estimates for $c_j$. This allows covariance-thresholded LARS to
estimate $c_j$ in a more stable way than the LARS.
It is clear that covariance-thresholded LARS presents an advantage
if population covariance is sparse. On the other hand, if the
covariance is non-sparse, covariance-thresholded LARS can still
outperform the LARS when the sample size is small or the data are
noisy.  This is because parsimonious estimates $(\hat{c}_\nu)_j$
of $c_j$ can be more robust against random variability of the
data.

Moreover, consider computing the direction of the solution paths
in Step 5, which is used for updating $(\hat{c}_\nu)_j$. LARS for
the lasso updates new directions with
$(\hat{\mathbf{\Sigma}})^{-1}_\mathcal{A}
sgn(\beta_{\mathcal{A}})$, whereas covariance-thresholded LARS
uses $\mathbf{\gamma}_\mathcal{A} =
(\hat{\mathbf{\Sigma}}_\nu)^{-1}_\mathcal{A}
sgn(\beta_{\mathcal{A}})$. Apparently, covariance-thresholded LARS
can exploit potential covariance sparsity to improve and stabilize
estimates of the directions of the solution paths. In addition,
the LARS for the lasso can stop early before all true variables
$S$ can be considered if $\hat{\mathbf{\Sigma}}_\mathcal{A}$ is
rank deficient at an early stage when sample size is limited.
Covariance-thresholding can mitigate this problem by proceeding further with properly chosen values of $\nu$. For example, when $\nu \to 1$,
$\hat{\mathbf{\Sigma}}_\nu$ converges towards the identity matrix
$\mathbf{I}$, which is full-ranked. 

\par \bigskip

\setcounter{chapter}{3}
\noindent {\bf 3. Theoretical Results on Variable Selection}
\par \smallskip
In this section, we derive sufficient conditions for
covariance-thresholded lasso to be consistent in selecting the
true variables. We relate covariance sparsity with variable
selection and demonstrate the pivotal role that covariance
sparsity plays in improving variable selection under
high-dimensionality. Furthermore, variable selection results for
the lasso under the random design are derived and compared with
those of the covariance-thresholded lasso. We show that the
covariance-thresholded lasso, by utilizing covariance sparsity
through a properly chosen thresholding level $\nu$, can improve
upon the lasso in terms of variable selection.

For simplicity, we assume that a solution for (\ref{def:CT-Lasso}) exists and denote the covariance-thresholded lasso
estimate $\hat \beta^{CT-Lasso}(\nu, \lambda_n)$ by $\hat
\beta^\nu$ in this section. Further, we let $supp(\beta) = \{j:
\beta_j \ne 0 \}$ represent the collection of indices of nonzero
coefficients. We say that the covariance-thresholded lasso
estimate $\hat \beta^\nu$ is variable selection consistent if $P(
supp(\hat \beta^\nu) = supp(\beta^*)) \to 1,\, \textrm{as }n\to
\infty$. In addition, we say that $\hat \beta^\nu$ is sign
consistent if $ P(sgn(\hat \beta^\nu) = sgn(\beta^*)) \to 1, \,
\textrm{as }n \to \infty$, where $sgn(t) = -1, 0, 1$ when $t <0$,
$t=0$ and $t>0$, respectively \citep*{Zhao06}. Obviously, sign
consistency is a stronger property and implies variable selection
consistency.

We introduce two quantities to characterize the sparsity of
$\mathbf{\Sigma}$ that plays a pivotal role in the performance of
covariance-thresholded lasso. Recall that $S$ and $C$ are
collections of the true and irrelevant variables, respectively.
Define
\begin{equation} \label{def:d^*}
d^*_{SS} = \max_{i\in S} \sum_{j \in S} 1(\sigma_{ij} \ne 0 )
\qquad \mbox{and}  \qquad d^*_{CS} = \max_{i \in C} \sum_{j \in S}
1(\sigma_{ij} \ne 0 ).
\end{equation}
$d^*_{SS}$ ranges between $1$ and $s$. When $d^*_{SS}=1$, all
pairs of the true variables are orthogonal. When $d^*_{SS}=s$,
there are at least one variable correlated with all other
variables.
Similarly, $d^*_{CS}$ is between $0$ and $s$. When $d^*_{CS}=0$,
the true and irrelevant variables are orthogonal to each other,
and, when $d^*_{CS}=s$, some irrelevant variables are correlated
with all the true variables. The values of $d^*_{SS}$ and
$d^*_{CS}$ represent the sparsity of covariance sub-matrices for
the true variables and between the irrelevant and true variables,
respectively. We have not specified the sparsity of the sub-matrix
for the irrelevant variables themselves. It will be clear later
that it is the structure of $\mathbf{\Sigma}_{SS}$ and
$\mathbf{\Sigma}_{CS}$ instead of $\mathbf{\Sigma}_{CC}$ that
plays the pivotal role in variable selection. We note that
$d^*_{SS}$ and $d^*_{CS}$ are related to another notion of
sparsity used in \citet*{Bickel08} to define the class of matrices
$\{\Sigma: \sigma_{ii}\le M, \sum_{j=1}^p 1(\sigma_{ij} \ne 0) \le
c_0(p) \mbox{ for } 1\le i \le p\}$, for $M$ given and $c_0(p)$ a
constant depending on $p$. We use the specific quantities
$d^*_{SS}$ and $d^*_{CS}$ in (\ref{def:d^*}) in order to provide
easier presentation of our results for variable selection. Our
results in this section can be applied to more general
characterizations of sparsity, such as in \citet*{Bickel08}.

In this paper, we employ two different types of matrix norms. For
an arbitrary matrix $A=[A_{ij}]$, the infinity norm is defined as
$\|A\|_\infty = \max_{i} \sum_{j} |A_{ij}|$, and the spectral norm
is defined as $\|A\| = \max_{x: \|x\| < 1}
\|Ax\|=\Lambda_{max}(A)$. We use $\Lambda_{max}(A)$ and
$\Lambda_{min}(A)$ to represent, respectively, the largest and
smallest eigenvalues of $A$.

\par \bigskip

\noindent {\bf 3.1. Sign Consistency of Covariance-thresholded
Lasso}
\par \smallskip
We develop sign consistency results for covariance-thresholded
lasso. Proofs for the results are presented in the Appendix.

We first provide conditions for the covariance-thresholded lasso
estimate $\hat{\beta}^\nu$ to have the same signs as the true
coefficients $\beta^*$ under the fixed design assumption.
Let $\bar{\rho} = \max_{j \in S} |\beta^*_j|$ and
$\underline{\rho} = \min_{j \in S} |\beta^*_j|$.

\begin{lemma} \label{Lemma:KKT.et}
Suppose that the data matrix $\mathbf{X}$ is fixed and $\nu$ is
given. Then, $sgn(\hat \beta^\nu) = sgn(\beta^*)$ if
\begin{equation}\label{Nonsingular}
\Lambda_{min}\left(\mathbf{\hat \Sigma}_{SS}^\nu \right) > 0,
\end{equation}
\begin{equation} \label{Ineq:et.KKT.1}
\left\|\mathbf{\hat \Sigma}_{CS}^\nu (\mathbf{\hat
\Sigma}_{SS}^\nu)^{-1} \right\|_\infty \left( \left\|{1\over n}
\mathbf{X}_S^T \epsilon\right\|_\infty +  s \nu \bar \rho +
\lambda_n
 \right) + s \nu \bar \rho + \left\|{1\over n} \mathbf{X}_{C}^T \epsilon
\right\|_\infty \le \lambda_n,
\end{equation}
and
\begin{equation} \label{Ineq:et.KKT.2}
\left\| (\mathbf{\hat \Sigma}_{SS}^\nu)^{-1} \right\|_\infty
\left( \left\|{1\over n} \mathbf{X}_S^T \epsilon\right\|_\infty +
s \nu \bar \rho + \lambda_n \right) < \underline{\rho}.
\end{equation}
\end{lemma}
The above (\ref{Nonsingular}), (\ref{Ineq:et.KKT.1}), and
(\ref{Ineq:et.KKT.2}) are derived from the Karush-Kuhn-Tucker
(KKT) conditions for the optimization problem presented in
(\ref{def:CT-Lasso}) when the solution, which may be a local minimum, exists. Following the arguments in \citet*{Zhao06}
and \citet*{Wainwright06}, these conditions are almost necessary
for $\hat \beta^\nu$ to have the correct signs. The condition
(\ref{Nonsingular}) is needed for (\ref{Ineq:et.KKT.1}) and
(\ref{Ineq:et.KKT.2}) to be valid. That is, the conditions
(\ref{Ineq:et.KKT.1}) and (\ref{Ineq:et.KKT.2}) are ill-defined if
$\mathbf{\hat \Sigma}_{SS}^\nu$ is singular.

Assume the random design setting so that $\mathbf{X}$ is drawn
from some distribution with population covariance
$\mathbf{\Sigma}$. We demonstrate
how the sparsity of $\mathbf{\Sigma}_{SS}$ and the procedure of
covariance-thresholding work together to ensure that the condition
$(\ref{Nonsingular})$ is satisfied. We impose the following moment
conditions on the random predictors $X_1, \ldots, X_p$:
\begin{equation} \label{Cond:X moments}
E X_j = 0, \qquad E X_j^{2d} \le d! M^d, \qquad 1\le j \le p,
\end{equation}
for some constant $M > 0$ and $d \in \mathbb{N}$. Assume that
\begin{equation} \label{Cond:Sigma1}
\Lambda_{min} \left( \mathbf{\Sigma}_{SS} \right)> 0
\end{equation}
and $d_{SS}^*$, $s$, and $n$ satisfy
\begin{equation} \label{cond:d*.1}
d_{SS}^* \sqrt{\log s} / \sqrt{n}  \to 0.
\end{equation}
We have the following lemma.
\begin{lemma} \label{Lemma:nonsingular.et}
Let $\nu = C \sqrt{\log s} / \sqrt{n}$ for some constant $C>0$.
Under the conditions (\ref{Cond:X moments}), (\ref{Cond:Sigma1}),
and (\ref{cond:d*.1}),
\begin{equation}\label{singular-conv-speed}
P\left( \Lambda_{min} \left( \mathbf{\hat \Sigma}_{SS}^\nu \right)
> 0 \right) \to 1.
\end{equation}
\end{lemma}
The rate of convergence for (\ref{singular-conv-speed}) depends on
the rate of convergence for (\ref{cond:d*.1}).
It is clear that the smaller $d^*_{SS}$ (or the sparser
$\mathbf{\Sigma}_{SS}$) is, the faster (\ref{cond:d*.1}), as well
as (\ref{singular-conv-speed}), converges. Equivalently, for
sample size $n$ fixed, the smaller $d^*_{SS}$ is, the larger the
probability that $\Lambda_{min} (\mathbf{\hat \Sigma}_{SS}^\nu)>
0$. In other words, covariance-thresholding can help to fix
potential rank deficiency of $\mathbf{\hat\Sigma}_{SS}$ when
$\mathbf{\Sigma}_{SS}$ is sparse.
In the special case when $\mathbf{\Sigma}_{SS} = I_p$ and
$d^*_{SS} = 1$, it can be shown that $ \mathbf{\hat
\Sigma}_{SS}^\nu$ is asymptotically positive definite provided
that $s = o(\exp(n))$.

Next, we investigate the remaining two conditions
(\ref{Ineq:et.KKT.1}) and (\ref{Ineq:et.KKT.2}) in Lemma 3.1. For
(\ref{Ineq:et.KKT.1}) and (\ref{Ineq:et.KKT.2}) to hold with
probability going to 1, additional assumptions including the
irrepresentable condition need to be imposed. Since the data
matrix $\mathbf{X}$ is assumed to be random, the original
irrepresentable condition needs to be stated in terms of the
population covariance matrix $\mathbf{\Sigma}$ as follows,
\begin{equation} \label{Cond:Sigma2}
\left\|\mathbf{\Sigma}_{C S}
(\mathbf{\Sigma}_{SS})^{-1}\right\|_\infty \le 1 - \epsilon,
\end{equation}
for some $0<\epsilon <1$. We note that the original
irrepresentable condition in \citet*{Zhao06} also involves the
signs of $\beta^*_{S}$.
To simplify presentation, we use the stronger condition
(\ref{Cond:Sigma2}) instead.
Obviously, (\ref{Cond:Sigma2}) does not directly imply that
$\|\mathbf{\hat \Sigma}_{CS}^\nu (\mathbf{\hat
\Sigma}_{SS}^\nu)^{-1} \|_\infty \le 1-\epsilon$. The next lemma
establishes the asymptotic behaviors of $\| (\mathbf{\hat
\Sigma}_{SS}^\nu)^{-1} \|_\infty$ and $\|\mathbf{\hat
\Sigma}_{CS}^\nu (\mathbf{\hat \Sigma}_{SS}^\nu)^{-1} \|_\infty$.
Let $\bar D = \|(\mathbf{\Sigma}_{SS})^{-1}\|_\infty$. Assume
\begin{equation} \label{cond:d*.2}
\bar D d^*_{SS} \sqrt{\log (p-s)} / \sqrt{n}  \to 0,
\end{equation}
\begin{equation} \label{Cond:(d*np)}
\bar D^2 d^*_{CS} d^*_{SS} \sqrt{\log (p-s)} / \sqrt{n}  \to 0.
\end{equation}

\begin{lemma} \label{Lemma:Irrep.et}
Suppose that $p-s>s$ and $\nu = C \sqrt{\log (s(p-s))} / \sqrt{n}$
for some constant $C>0$. Under conditions (\ref{Cond:X moments}),
(\ref{Cond:Sigma1}), (\ref{Cond:Sigma2}), (\ref{cond:d*.2}), and
(\ref{Cond:(d*np)}),
\begin{equation} \label{8.1}
P \left( \left\| \left(\hat{\mathbf{\Sigma}}_{SS}^\nu \right)^{-1}
\right\|_{\infty} \le \bar{D} \right) \to 1,
\end{equation}
\begin{equation} \label{Irrep1}
P\left(  \left\|\mathbf{\hat \Sigma}_{CS}^\nu (\mathbf{\hat
\Sigma}_{SS}^\nu)^{-1}\right\|_\infty \le 1 - {\epsilon \over 2}
\right) \to 1.
\end{equation}
\end{lemma}
The above lemma indicates that with a properly chosen thresholding
parameter $\nu$ and sample size depending on covariance-sparsity
quantities $d^*_{SS}$ and $d^*_{CS}$, both $\| (\mathbf{\hat
\Sigma}_{SS}^\nu)^{-1} \|_\infty$ and $\|\mathbf{\hat
\Sigma}_{CS}^\nu (\mathbf{\hat \Sigma}_{SS}^\nu)^{-1} \|_\infty$
behave as their population counterparts
$\|(\mathbf{\Sigma}_{SS})^{-1}\|_\infty$ and
$\|\mathbf{\Sigma}_{CS} \mathbf{\Sigma}_{SS}^{-1}\|_\infty$,
asymptotically. Again, the influence of the sparsity of
$\mathbf{\Sigma}$ on $\| (\mathbf{\hat \Sigma}_{SS}^\nu)^{-1}
\|_\infty$ and $\|\mathbf{\hat \Sigma}_{CS}^\nu (\mathbf{\hat
\Sigma}_{SS}^\nu)^{-1} \|_\infty$ is shown through $d^*_{CS}$ and
$d^*_{SS}$. Asymptotically, the smaller $d^*_{CS}$ and $d^*_{SS}$
are, the faster (\ref{8.1}) and (\ref{Irrep1}) converge. Or
equivalently, for sample size $n$ fixed, the smaller $d^*_{CS}$
and $d^*_{SS}$ are, the larger the probabilities in (\ref{8.1})
and (\ref{Irrep1}) are. In the special case when $d^*_{CS}=0$ or
$\mathbf{\Sigma}_{CS}$ is a zero matrix, condition
(\ref{Cond:(d*np)}) is always satisfied.

Finally, we are ready to state the sign consistency result for
$\hat \beta^\nu$. With the help of Lemmas 1--3 stated above, the
only issue left is to show the existence of a proper $\lambda_n$
such that (\ref{Ineq:et.KKT.1}) and (\ref{Ineq:et.KKT.2}) hold
with probability going to $1$. One more condition is needed. We
assume that
\begin{equation} \label{Cond:dim.1}
\bar D  \bar \rho s \sqrt{\log (p-s)} / (\underline{\rho} \sqrt{n}
) \to 0.
\end{equation}

\begin{thm} \label{Thm:et.vsc}
Suppose that $p-s>s$, $\nu =  C \sqrt{\log(s(p-s))} / \sqrt{n}$
for some constant $C>0$, and $\lambda_n$ is chosen such that
$\lambda_n \to 0$,
\begin{equation} \label{cond:lambda.et}
\sqrt{n} \lambda_n / ( s\overline{\rho} \sqrt{\log (p-s)} ) \to
\infty, \qquad \mbox{and} \qquad \bar{D} \lambda_n /
\underline{\rho} \to 0.
\end{equation}
Then, under conditions (\ref{Cond:X moments}),
(\ref{Cond:Sigma1}), (\ref{Cond:Sigma2}), (\ref{Cond:(d*np)}), and
(\ref{Cond:dim.1}),
\begin{equation} \label{probability}
P \left(sgn(\hat \beta^\nu) = sgn(\beta^*)
\right) \to 1.
\end{equation}
\end{thm}
We note that the assumption $p-s>s$ is natural for
high-dimensional sparse models, which usually have a large number
of irrelevant variables. This assumption effects the conditions
(\ref{Cond:(d*np)}) and (\ref{Cond:dim.1}) as well as choices of
$\nu$ and $\lambda_n$. When $p-s<s$, that is a non-sparse linear
model is assumed, the conditions for $\hat{\beta}^\nu$ to be sign
consistent need to be modified by choosing $\nu$ as $\nu = C
\sqrt{ \log s} / \sqrt{n}$ and replacing $\sqrt{\log (p-s)}$ by
$\sqrt{\log s}$ in conditions (\ref{Cond:(d*np)}),
(\ref{Cond:dim.1}), and (\ref{cond:lambda.et}).

It is possible to establish the convergence rate for the
probability in (\ref{probability}) more explicitly. For simplicity
of presentation, we provide such a result under a special case in
the following theorem.

\begin{thm} \label{thm.probrate}
Suppose that conditions (\ref{Cond:X moments}),
(\ref{Cond:Sigma1}), and (\ref{Cond:Sigma2}) hold and $\bar{D}$,
$\underline{\rho}$, and $\bar{\rho}$ are constants. Let $\lambda_n
= n^{-c}$, $\nu = n^{-c_1}$, $d_S^* = \max\{d^*_{SS}, d^*_{CS} \}
= n^{c_2}$, $s = n^{c_3}$, and $\log p = o(n^{1-2c} +
n(n^{-c_2}-n^{-c_1})^2)$, where $c$, $c_1$, $c_2$, and $c_3$ are
positive constants such that $c < 1/2$, $c_1 < 1/2$, $c_2 < 1/4$,
$c_2 < c_3$, and $c_3+c < c_1$. Then,
\begin{equation} \label{probrate}
P \left( sgn(\hat \beta^\nu ) = sgn(\beta^*) \right) \ge 1 -
O\left( \exp(-\alpha_1 n^{1-2c}) \right) - O \left( \exp(-\alpha_2
n(n^{-2c_2} -n^{-c_1})^2 ) \right) \to 1,
\end{equation}
where $\alpha_1$ and $\alpha_2$ are some positive constants
depending on $\epsilon$, $\bar D$, $M$ and $\underline{\rho}$.
\end{thm}

The proof of Theorem \ref{thm.probrate}, which we omit, is similar
to that of Theorem \ref{Thm:et.vsc}. We note that the conditions
on dimension parameters in Theorem \ref{Thm:et.vsc} are now
expressed in the convergence rate of (\ref{probrate}). It is clear
that the smaller $d_S^*$ is, the larger the probability is in
(\ref{probrate}).

\par \bigskip

\noindent {\bf 3.2. Comparison with the Lasso}
\par \smallskip
We compare sign consistency results of covariance-thresholded
lasso with those of the lasso. By choosing $\nu=0$, the
covariance-thresholded lasso estimate $\hat \beta^\nu$ can be
reduced to the lasso estimate $\hat \beta^0$. Results on sign
consistency of the lasso have been established in the literature
(\citet*{Zhao06}, \citet*{Meinshausen06}, \citet*{Wainwright06}).
To facilitate comparison, we restate sign consistency results for
$\hat \beta^0$ in the same way that we presented results for $\hat
\beta^\nu$ in Section 3.1
. The proofs, which we omit, for sign consistency of $\hat
\beta^0$ is similar to those for $\hat \beta^\nu$.

First, assuming fixed design, we have the sufficient and almost
necessary conditions for $sgn(\hat \beta^{0}) = sgn(\beta^*)$ as
in (\ref{Nonsingular})-(\ref{Ineq:et.KKT.2}) with $\nu = 0$.

Next, we assume the random design. Analogous to Lemma
{\ref{Lemma:nonsingular.et}}, the sufficient conditions for $P(
\Lambda_{min} ( \mathbf{\hat \Sigma}_{SS} ) > 0) \to 1$ are
(\ref{Cond:X moments}), (\ref{Cond:Sigma1}), and
\begin{equation} \label{cond:s.1}
s \sqrt{\log s} / \sqrt{n} \to 0.
\end{equation}
Compared to (\ref{cond:d*.1}), (\ref{cond:s.1}) is clearly more
demanding since $d^*_{SS}$ is always less than or equal to $s$.
Note that a necessary condition for $\mathbf{\hat \Sigma}_{SS}$ to
be non-singular is $s \le n$, which is not required for
$\mathbf{\hat \Sigma}_{SS}^\nu$. Thus, the non-singularity of the
sample covariance sub-matrix $\mathbf{\hat \Sigma}_{SS}$ is harder
to attain than that of $\mathbf{\hat \Sigma}_{SS}^\nu$.
In other words, covariance-thresholded lasso may increase the rank
of $\mathbf{\hat\Sigma}_{SS}$ by thresholding. When
$\mathbf{\Sigma}$ is sparse, this can be beneficial for variable
selection under the large $p$ and small $n$ scenario.

To ensure that $P ( \| (\hat{\mathbf{\Sigma}}_{SS} )^{-1}
\|_{\infty} \le \bar{D} ) \to 1$ and $P( \|\mathbf{\hat
\Sigma}_{CS} (\mathbf{\hat \Sigma}_{SS})^{-1}\|_\infty \le 1 -
\epsilon / 2 ) \to 1$, as in Lemma \ref{Lemma:Irrep.et} with $\nu
= 0$, we further assume the conditions (\ref{Cond:Sigma2}) and
\begin{equation} \label{Cond:(snp)}
\bar D^2 s^2 \sqrt{\log (p-s)} / \sqrt{n}  \to 0.
\end{equation}
We note that (\ref{Cond:(snp)}) is the main condition that
guarantees that $\mathbf{\hat \Sigma}$ satisfies the
irrepresentable condition with probability going to 1. Compared
with (\ref{Cond:(d*np)}), (\ref{Cond:(snp)}) is clearly more
demanding since $s$ is larger than both $d^*_{CS}$ and $d^*_{SS}$.
This implies that it is harder for $\hat{\mathbf{\Sigma}}$ than
for $\hat{\mathbf{\Sigma}}_\nu$ to satisfy the irrepresentable
condition.
In other words, covariance-thresholded lasso is more likely to be
variable selection consistent than the lasso when data are
randomly generated from a distribution that satisfies
(\ref{Cond:Sigma2}).

Finally, with the additional condition,
\begin{equation} \label{Cond:dim.2}
\bar D \sqrt{\log s} / (\underline{\rho} \sqrt{n} ) \to 0,
\end{equation}
we arrive at the sign consistency of the lasso as the following.

\begin{cor} \label{cor:lasso.vsc}
Assume that the conditions (\ref{Cond:X moments}),
(\ref{Cond:Sigma1}), (\ref{Cond:Sigma2}), (\ref{Cond:(snp)}), and
(\ref{Cond:dim.2}) are satisfied. If $\lambda_n$ is chosen such
that $\lambda_n \to 0$,
\begin{equation} \label{Cond:lambda.lasso}
 \sqrt{n} \lambda_n / \sqrt{\log(p-s)} \to \infty, \qquad
\textrm{and} \qquad \bar{D} \lambda_n  / \underline{\rho} \to 0,
\end{equation}
then, $P \left(sgn(\hat \beta^{0}) = sgn(\beta^*) \right) \to 1$.

\end{cor}
Compare Corollary \ref{cor:lasso.vsc} with Theorem
\ref{Thm:et.vsc} for covariance-thresholded lasso. We see that
conditions (\ref{Cond:X moments}), (\ref{Cond:Sigma1}),
(\ref{Cond:Sigma2}) on random predictors, in particular the
covariances, are the same, but conditions on dimension parameters,
such as $n$, $p$, $s$, etc., are different. When the population
covariance matrix $\mathbf{\Sigma}$ is sparse, condition
(\ref{Cond:(d*np)}) on dimension parameters is much weaker for
covariance-thresholded lasso than condition (\ref{Cond:(snp)}) for
the lasso .
This shows that covariance-thresholded lasso can improve the
possibility of there existing a consistent solution.
However, a trade-off presents in the selection of tuning
parameters $\lambda_n$. The first condition in
(\ref{cond:lambda.et}) for covariance-thresholded lasso is clearly
more restricted than the condition in (\ref{Cond:lambda.lasso})
for the lasso. This results in a more restricted range for valid
$\lambda_n$. We argue that compared with the existence of
consistent solution, the range of the $\lambda_n$ is of secondary
concern.

We note that a related sign consistency result under random design
for the lasso has been established in \citet*{Wainwright06}. They
assume that the predictors are normally distributed and utilize
the resulting distribution of the sample covariance matrix. The
conditions used in \citet*{Wainwright06} include
(\ref{Cond:Sigma1}), (\ref{Cond:Sigma2}),
$\Lambda_{max}(\mathbf{\Sigma}) < \infty$, $\bar{D} < \infty$, $
\log (p-s) / (n-s) \to 0$, $\sqrt{\log s} / (\underline{\rho}
\sqrt{n}) \to 0$, and $ n > 2 \left(\Lambda_{max}
(\mathbf{\Sigma})/(\epsilon^2 \Lambda_{min}
(\mathbf{\Sigma}_{SS})) + \nu \right)s \log(p-s) + s + 1$, for
some constant $\nu > 0$. In comparison, we assume, in this paper,
that the random predictors follow the more general moment
conditions (\ref{Cond:X moments}), which contain the Gaussian
assumption as a special case. Moreover, we use a new approach to
establish sign consistency that can incorporate the sparsity of
the covariance matrix.

\par \bigskip

\setcounter{chapter}{4}
\noindent {\bf 4. Simulations}
\par \smallskip
In this section, we examine the finite-sample performances of the
covariance-thresholded lasso for $p \ge n$ and compare them to
those of the lasso, adaptive lasso with univariate as initial
estimates, UST, scout(1,1), scout(2,1), and elastic net.  Further, we propose a novel variant
of cross-validation that allows improved variable selection when
$n$ is much less than $p$. We note that the scout(1,1) procedure can be computationally expensive.  Results for scout(1,1) that take longer than 5 days on an RCAC cluster were not shown.

We compare variable selection performances using the $G$-measure,
$G=\sqrt{sensitivity*specificity}$. $G$ is defined as the geometric mean between sensitivity, $\textrm{(no. of true positives)}/s$, and specificity, $1-(\textrm{no. of false positives})/(p-s)$. Sensitivity and specificity can be interpreted as the proportion of selecting the true variables correctly and discarding the irrelevant variables correctly, respectively. Sensitivity can also be defined as 1 minus false negative rate and specificity as 1 minus false positive rate.
A value close to 1 for $G$ indicates
good selection, whereas a value close to 0 implies that few true
variables or too many irrelevant variables are selected, or both.
Furthermore, we compare prediction accuracy using the relative
prediction error (RPE), $\text{RPE}=(\hat{\beta} - \beta^*)^T
\mathbf{\Sigma} (\hat{\beta} - \beta^*)/\sigma^2$ where
$\mathbf{\Sigma}$ is the population covariance matrix.  The RPE is
obtained by re-scaling the mean-squared error (ME), as in
\citet{Tibshirani96}, by $1/\sigma^2$.

We first present variable selection results using best-possible
selection of tuning parameters, where tuning parameters are
selected ex post facto based on the best $G$. This procedure is
useful in examining variable selection performances, free from
both inherent variabilities in estimating the tuning parameters
and possible differences in the validation procedures used.
Moreover, it is important as an informant of the possible
potentials of the methods examined. We present median G out of 200
replications using best-possible selection of tuning parameters.
Standard errors based on 500 bootstrapped re-samplings are very
small, in the hundredth decimal place, for median G and are not
shown.

Results from best-possible selection of tuning parameters allow us to understand the potential advantages of the methods if one chooses their tuning parameters correctly.
However, in practice, possible errors due to the selection of tuning
parameters may sometimes overcome the benefit of introducing
them. Hence, we include additional results that use
cross-validation to select tuning parameters.

We study variable selection methods using a novel variant of the
usual cross-validation to estimate the model complexity parameter
$\lambda_n$ that allows improved variable selection when $p \gg
n$. Conventional cross-validation selects tuning parameters based
upon the minimum validation error, obtained from the average of
sum-of-squares errors from each fold. It is well known
that, when the sample size $n$ is large compared with the number
of predictors $p$, procedures such as cross-validation that are
prediction-based tend to over-select. This is because, when the
sample size is large, regression methods tend to produce small but
non-zero estimates for coefficients of irrelevant variables and
over-training occurs. On the other hand, we note that a different
scenario occurs when $p \gg n$. In this situation,
prediction-based procedures, such as the usual cross-validation,
tend to under-select important variables. This is because, when
$n$ is small, inclusion of a relatively few irrelevant variables
can increase the validation error dramatically, resulting in
severe instability and under-representation of important
variables.  In this paper, we propose to use a variant of the
usual cross-validation, in which we include additional variables
by decreasing $\hat{\lambda}_n$ for up to 1 standard deviation of
the validation error at the minimum. Through extensive empirical
studies, we found that this strategy often works well to
prevent under-selection when $n/\sqrt{p}<5$, which corresponds to
$n<50$ when $p=100$ and $n<224$ when $p=2000$. For $n/\sqrt{p} \ge
5$ and sample size $n$ only moderately large, we use the usual
cross-validation at the minimum. We note that \citet*[p.
216]{Hastie01} have described a related strategy that discards
variables up to 1 standard deviation of the minimum
cross-validation error for use when $n$ is large relative to $p$
and over-selection is severe. In Table \ref{tab:ex1}-\ref{tab:ex3}, we present median RPE, number of true and false positives, sensitivity, specificity, and G out of 200 replications using modified
cross-validation for selecting tuning parameters. The smallest 3 values of median RPE and largest 3 of median G are highlighted in bold. Standard errors
based on 500 bootstrapped re-samplings are further reported in parentheses for median RPE and G.  In Table \ref{tab:cv}, we provide an additional simulation
study to illustrate the modified cross-validation.

In each example, we simulate 200 data sets from the true model,
$\mathbf{y}=\mathbf{X} \beta^* + \sigma \epsilon$, where $\epsilon \sim
N(0,I)$.  $\mathbf{X}$ is generated each time from  $N(0,\mathbf{\Sigma})$, and we vary $\mathbf{\Sigma}$, $\beta^*$, and $\sigma$ in each example to illustrate performances across a variety of situations. We choose the tuning parameter $\gamma$ from $\{0, 0.5, 1, 2\}$ for both adaptive lasso \citep*{Zou06} and
covariance-thresholded lasso with adaptive thresholding.  The adaptive lasso seeks to improve upon the lasso by applying the weights $1/|\hat{\mathbf{\beta}}_0|^\gamma$, where $\hat{\mathbf{\beta}}_0$ is an initial estimate, in order to penalize each coefficient differently in the $L1$-norm of the lasso.  The larger $\gamma$ is the less the shrinkage applied to coefficients of large magnitudes.  The candidate values used for $\gamma$ are suggested in \citet*{Zou06} and found to work well in practice.

\example{1} (Autocorrelated.) This example has $p=100$ predictors
with coefficients $\beta^*_j=3$ for $j\in\{1,\ldots,5\}$,
$\beta^*_j=1.5$ for $j\in\{11,\ldots,15\}$, and $\beta^*_j=0$
otherwise. $\mathbf{\Sigma}_{ij}=0.5^{|i-j|}$ for all $i,j$, and
$\sigma=9$. Signal-to-noise ratio (SNR) ${\beta^*}^T
\mathbf{\Sigma} \beta^* / \sigma^2$ is approximately $1.55$. This
example, similar to Example 1 in \citep*{Tibshirani96}, has an
approximately sparse covariance structure, as elements away from
the diagonal can be extremely small.

\begin{table}[!tb]\renewcommand{\baselinestretch}{1}
\caption{\label{tab:ex1} Example 1 performance results using
fivefold cross-validation based on 200 replications. 
} \centering
\begin{small}
\begin{tabular}{|l|l||l|cc|lll|}
\hline \emph{n} & \emph{Method} & \textit{rpe} & \textit{TP} & \textit{FP} & \textit{sens} &
\textit{spec} & \textit{$G$} \\ \hline
20 & Lasso              &\textbf{1.284} (0.043) & 4.0 & 13.0 &0.40 &0.86 &0.577 (0.003)\\
& Adapt Lasso        &\textbf{1.301} (0.060) & 4.0 & 12.0 &0.40 &0.87 &0.581 (0.006)\\
& UST                &3.001 (0.223) &7.0 &28.0  & 0.70 & 0.69 & \textbf{0.690} (0.008) \\
& Scout(1,1)         &\textbf{1.164} (0.027) &10.0  &90.0  & 1.00 & 0.00 & 0.000 (0.000) \\
& Scout(2,1)         &1.474 (0.053) &6.0  &39.0  & 0.60 & 0.57 & 0.474 (0.023) \\
& Elastic net        &1.630 (0.097) &7.0  &31.0  &0.70 &0.63 &0.633 (0.021)\\
& CT-Lasso hard      &1.713 (0.100) &5.0  &22.5  &0.60 &0.77 &0.593 (0.013)\\
& CT-Lasso soft      &1.586 (0.051) &6.0  &20.5  &0.60 &0.78 & \textbf{0.667} (0.007)\\
& CT-Lasso adapt     &1.602 (0.055) &6.0  &20.0  &0.60 &0.78 &\textbf{0.654} (0.006)\\
\hline
40 & Lasso               &1.095 (0.052)  &6.0  &27.0  &0.60   &0.71   &0.672 (0.003)\\
& Adapt Lasso         &\textbf{1.047} (0.038)  &7.0  &21.0  &0.70   &0.77   &0.706 (0.007)\\
& UST                 &1.918 (0.098) &8.0 &28.0  & 0.80 & 0.69 & \textbf{0.742} (0.006) \\
& Scout(1,1)          &\textbf{0.814} (0.016)  &10.0  &90.0  & 1.00 & 0.00 & 0.000 (0.025) \\
& Scout(2,1)          &1.125 (0.039)  &9.0  &53.0  & 0.90 & 0.41 & 0.544 (0.029) \\
& Elastic net         &1.490 (0.066)  &8.0  &32.0  &0.90   &0.63   &0.683 (0.010)\\
& CT-Lasso hard       &1.221 (0.072)  &7.0  &23.0  &0.70   &0.74   &0.704 (0.008)\\
& CT-Lasso soft       &1.068 (0.055)  &7.0  &23.0  &0.80   &0.77   &\textbf{0.739} (0.007)\\
& CT-Lasso adapt      &\textbf{1.063} (0.045)  &7.0  &23.0  &0.80   &0.78   &\textbf{0.743} (0.007)\\
\hline
80 & Lasso           &0.379 (0.010) &8.0  &19.0  &0.80    &0.79   &0.794 (0.005)\\
& Adapt Lasso     &0.367 (0.013) &8.0  &15.0  &0.80    &0.82   &0.800 (0.005)\\
& UST                 &0.360 (0.011) &8.0 &5.0  & 0.80 & 0.94 & \textbf{0.851} (0.012) \\
& Scout(1,1)          &\textbf{0.245} (0.007)  &8.0  &8.0  & 0.80 & 0.91 & \textbf{0.854} (0.008) \\
& Scout(2,1)          &0.399 (0.014)  &6.5  &7.0  & 0.65 & 0.92 & 0.762 (0.006) \\
& Elastic net     &\textbf{0.307} (0.014) &9.0  &10.0  &0.90    &0.90   &\textbf{0.866} (0.006)\\
& CT-Lasso hard   &0.349 (0.013) &8.0  &8.0  &0.80    &0.94   &0.795 (0.010)\\
& CT-Lasso soft   &\textbf{0.284} (0.011) &8.0  &6.5  &0.80    &0.94   &0.827 (0.008)\\
& CT-Lasso adapt  &0.316 (0.017) &8.0  &8.0  &0.80    &0.93   &0.823 (0.009)\\
\hline
\end{tabular}
\end{small}
\end{table}
Figure \ref{fig:best}(a) depicts variable selection results using
best-possible selection of tuning parameters. We see that the
covariance-thresholded lasso methods dominate the lasso,
adaptive lasso, and UST in terms of variable selection for $p \ge n$. The
performances of lasso and adaptive lasso deteriorate precipitously
as $n$ becomes small, whereas those of the covariance-thresholded
lasso methods decrease at a relatively slow pace.  Furthermore,
the covariance-thresholded lasso methods dominate the elastic net and scout for $n$ small.  We also observe that the scout procedures and elastic net perform very similarly.  This is not surprising as \citet{Witten08} have shown in Section 2.5.1 of their paper that scout(2,1), by regularizing the inverse covariance matrix, is very similar to the elastic net.

Results from best-possible selection provide information on the
potentials of the methods examined. In Table \ref{tab:ex1}, we
present results using cross-validation to illustrate performances
in practice. The covariance-thresholded lasso methods tend to dominate
the lasso, adaptive lasso, scout, and elastic net in terms of variable
selection for $n$ small. The UST presents good variable selection performances but large prediction errors. We note that, due to its large bias, the UST cannot be legitimately applied with cross-validation that uses validation error to select tuning parameters, especially when the coefficients are disparate and some correlations are large.  The advantages of covariance-thresholded
lasso with hard thresholding is less apparent compared with those
of soft and adaptive thresholding. This suggests that continuous
thresholding of covariances may achieve better performances than
discontinuous ones using cross-validation. We note that the scout procedures perform surprisingly poorly compared with the covariance-thresholded lasso and the elastic net in terms of variable selection when $n$ is small.  As the scout and elastic net are quite similar in terms of their potentials for variable selection as shown in Figure \ref{fig:best}(a), the differences seem to come from the additional re-scaling step of the scout, where the scout re-scales its initial estimates by multiplying them with a scalar $\hat{c} = \arg\min_c \|\mathbf{y}-c\mathbf{X}\hat{\mathbf{\beta}}\|^2$.  This strategy can sometimes be useful in improving prediction accuracy.  However, when $n$ is small compared with $p$, standard deviations of validation errors for the scout can often be large, which may cause variable selection performances to suffer for cross-validation.  We additionally note that, when $p \gg n$ and SNR is low as in this example, high specificity can sometimes be more important for prediction accuracy than high sensitivity. This is because, when $n$ is small, coefficients of irrelevant variables can be given large estimates, and inclusion of but a few irrelevant variables can significantly deteriorate prediction accuracy. In Table \ref{tab:ex1}, we see that the lasso and adaptive lasso have good prediction accuracy for $n=20$ though it selects less than half of the true variables.

\begin{table}[!tb]\renewcommand{\baselinestretch}{1}
\caption{\label{tab:ex2} Example 2 performance results using
fivefold cross-validation based on 200 replications. 
} \centering
\begin{small}
\begin{tabular}{|l|l||l|cc|lll|}
\hline \emph{n} & \emph{Method} & \textit{rpe} & \textit{TP} & \textit{FP} & \textit{sens} &
\textit{spec} & \textit{$G$} \\ \hline
20 & Lasso          &0.341 (0.027)  &2.0    &9.0   &0.10 &0.89   &0.302 (0.009)\\
& Adapt Lasso       &0.352 (0.028)  &2.0    &9.0   &0.10 &0.89   &0.301 (0.006)\\
& Elastic net       &0.967 (0.137)  &14.0   &51.5  &0.70 &0.36   &\textbf{0.437} (0.011)\\
& UST               &28.930 (0.836) &19.0   &73.0  &0.95 &0.09   &0.296 (0.012) \\
& Scout(1,1)        &NA &NA &NA &NA &NA &NA \\
& Scout(2,1)        &\textbf{0.062} (0.004)  &20.0   &80.0  &1.00 &0.00   &0.000 (0.000) \\
& CT-Lasso hard     &0.383 (0.018)  &3.0    &11.0  &0.15 &0.86   &0.370 (0.013)\\
& CT-Lasso soft     &\textbf{0.231} (0.015)  &6.5    &23.0  &0.33 &0.71   &\textbf{0.465} (0.008)\\
& CT-Lasso adapt    &\textbf{0.302} (0.019)  &6.5    &23.0  &0.33 &0.71   &\textbf{0.461} (0.012)\\
\hline
40 & Lasso          &0.348  (0.017)  &5.0    &18.5  &0.25 &0.77   &0.429  (0.014)\\
& Adapt Lasso       &0.315  (0.024)  &5.0    &17.0  &0.25 &0.79   &0.417  (0.008)\\
& Elastic net       &0.739  (0.094)  &16.0   &58.0  &0.80 &0.28   &0.426  (0.014)\\
& UST               &26.189 (1.001)  &20.0   &77.0  &1.00 &0.04   &0.194  (0.007) \\
& Scout(1,1)        &NA &NA &NA &NA &NA &NA \\
& Scout(2,1)        &\textbf{0.043}  (0.004)  &20.0   &80.0  &1.00 &0.00   &0.000  (0.000) \\
& CT-Lasso hard     &0.363  (0.018)  &6.0    &21.0  &0.30 &0.74   &\textbf{0.450}  (0.008)\\
& CT-Lasso soft     &\textbf{0.269}  (0.023)  &10.0   &35.0  &0.50 &0.56   &\textbf{0.485}  (0.006)\\
& CT-Lasso adapt    &\textbf{0.306}  (0.029)  &8.0    &31.0  &0.40 &0.61   &\textbf{0.482}  (0.006)\\
\hline
80 & Lasso          &0.123  (0.004)  &5.0   &14.0  &0.25 &0.83   &0.440  (0.008)\\
& Adapt Lasso       &0.122  (0.004)  &4.0   &14.0  &0.20 &0.83   &0.423  (0.009)\\
& Elastic net       &\textbf{0.089}  (0.006)  &14.0  &48.5  &0.70 &0.39   &0.461  (0.012)\\
& UST               &\textbf{0.042}  (0.003)  &18.0  &66.0  &0.90 &0.18   &0.393  (0.006) \\
& Scout(1,1)        &NA &NA &NA &NA &NA &NA \\
& Scout(2,1)        &\textbf{0.038}  (0.002)  &20.0  &80.0  &1.00 &0.00   &0.000  (0.000) \\
& CT-Lasso hard     &0.159  (0.007)  &6.0   &17.5  &0.30 &0.78   &\textbf{0.468}  (0.008)\\
& CT-Lasso soft     &0.107  (0.009)  &9.0   &27.0  &0.45 &0.66   &\textbf{0.521}  (0.007)\\
& CT-Lasso adapt    &0.129  (0.014)  &8.0   &24.0  &0.40 &0.70   &\textbf{0.503}  (0.007)\\
\hline
\end{tabular}
\end{small}
\end{table}

\example{2} (Constant covariance.) This example has $p=100$
predictors with $\beta^*_j=3$ for $j\in\{11,\ldots,20\}$,
$\beta^*=1.5$ for $j\in\{31,\ldots,40\}$, and $\beta^*_j=0$
otherwise. $\mathbf{\Sigma}_{ij}=0.95$ for all $i$ and $j$ such
that $i \neq j$, and $\sigma=15$. SNR is approximately $8.58$.
This example, derived from Example 4 in \citet*{Tibshirani96},
presents an extreme situation where all non-diagonal elements of
the population covariance matrix are nonzero and constant.

In Figure \ref{fig:best}(b), we see that the
covariance-thresholded lasso methods dominate over the lasso and
adaptive lasso, especially for $n$ small. This example shows that
sparse covariance thresholding may still improve variable
selection when the underlying covariance matrix is non-sparse.  Furthermore, covariance-thresholded lasso methods with soft and adaptive thresholding perform better than that with hard thresholding.  
Interestingly, we see that the performance of UST decreases with increasing $n$ and drops below that of the lasso for $n \ge 30$.  This example demonstrates that the UST may not be a good general procedure for variable selection and can sometimes fail unexpectedly.  We note that this is a challenging example for variable
selection in general. By the irrepresentable condition \citep*{Zhao06}, the
lasso is not variable selection consistent under this scenario.
The median $G$ values in Figure \ref{fig:best}(b) usually increase much
slower with increasing $n$ in comparison with those of Example 1
in Figure \ref{fig:best}(a), even though SNR is higher.

Table \ref{tab:ex2} shows that the covariance-thresholded lasso
methods and the elastic net dominate over the lasso and adaptive
lasso in terms of variable selection when using cross-validation
to select tuning parameters.
The UST under-performs the lasso and adaptive lasso in terms of variable selection.  Scout(2,1) does the worst in terms of variable selection by including all variables but presents the best prediction error.  Again, we note that this may be due to the re-scaling step employed by the scout, which may sometimes improve performance in prediction but often suffers in terms of variable selection, especially when the sample size is small.

\begin{table}[!tb]\renewcommand{\baselinestretch}{1}
\caption{\label{tab:ex3} Example 3 performance results using
fivefold cross-validation based on 200 replications. 
} \centering
\begin{small}
\begin{tabular}{|l|l||l|cc|lll|}
\hline \emph{n} & \emph{Method} & \textit{rpe} & \textit{TP} & \textit{FP} & \textit{sens} &
\textit{spec} & \textit{$G$} \\ \hline
20 & Lasso           &\textbf{0.751} (0.024) &5.0  &12.0  &0.50 &0.87   &0.650 (0.003)\\
& Adapt Lasso        &\textbf{0.773} (0.035) &5.0  &11.0  &0.50 &0.88   &0.652 (0.005)\\
& UST                 &3.340 (0.202)  &8.0 &31.0  & 0.80 & 0.66 & \textbf{0.712} (0.007) \\
& Scout(1,1)          &\textbf{0.682} (0.022)  &10.0  &90.0  & 1.00 & 0.00 & 0.000 (0.000) \\
& Scout(2,1)          &1.274 (0.081)  &8.0  &37.5  & 0.80 & 0.58 & 0.536 (0.017) \\
& Elastic net        &1.891 (0.203) &9.0  &32.0  &0.90 &0.64   &0.660 (0.015)\\
& CT-Lasso hard      &1.542 (0.105) &6.5  &20.5  &0.70 &0.77   &0.636 (0.018)\\
& CT-Lasso soft      &1.240 (0.058) &7.0  &22.5  &0.70 &0.77   &\textbf{0.700} (0.013)\\
& CT-Lasso adapt     &1.427 (0.069) &7.0  &19.0  &0.70 &0.76   &\textbf{0.665} (0.015)\\
\hline
40 & Lasso        &0.729 (0.044) &8.0  &27.0  &0.80 &0.70   &0.748 (0.008)\\
& Adapt Lasso     &\textbf{0.659} (0.038) &8.0  &21.5  &0.80 &0.76   &0.789 (0.008)\\
& UST                 &1.784 (0.068) &10.0 &33.0  & 1.00 & 0.63 & 0.782 (0.010) \\
& Scout(1,1)          &\textbf{0.518} (0.013)  &10.0  &68.5  & 1.00 & 0.24 & 0.475 (0.103) \\
& Scout(2,1)          &\textbf{0.628} (0.037)  &10.0  &54.5  & 1.00 & 0.39 & 0.616 (0.028) \\
& Elastic net     &1.101 (0.057) &10.0  &35.0  &1.00 &0.62   &0.748 (0.012)\\
& CT-Lasso hard   &0.808 (0.045) &9.0  &21.0  &0.90 &0.76   &\textbf{0.806} (0.012)\\
& CT-Lasso soft   &0.723 (0.026) &9.0  &22.0  &0.90 &0.76   &\textbf{0.815} (0.007)\\
& CT-Lasso adapt  &0.760 (0.046) &9.0  &22.0  &0.90 &0.76   &\textbf{0.819} (0.009)\\
\hline
80 & Lasso        &0.221 (0.013) &9.0  &24.0  &0.90 &0.73   &0.825 (0.007)\\
& Adapt Lasso     &0.222 (0.017) &10.0  &19.0  &1.00 &0.79   &0.864 (0.006)\\
& UST             &0.104 (0.008) &10.0 &6.0  & 1.00 & 0.93 & \textbf{0.946} (0.004) \\
& Scout(1,1)      &0.070 (0.005)  &10.0  &5.5  & 1.00 & 0.94 & 0.937 (0.004) \\
& Scout(2,1)      &0.070 (0.003)  &10.0  &7.0  & 1.00 & 0.92 & \textbf{0.938} (0.004) \\
& Elastic net     &0.104 (0.009) &10.0  &9.0  &1.00 &0.90   &0.937 (0.005)\\
& CT-Lasso hard   &\textbf{0.069} (0.005) &9.0  &3.0  &0.90 &0.97   &\textbf{0.938} (0.004)\\
& CT-Lasso soft   &\textbf{0.063} (0.005) &10.0  &4.0  &1.00 &0.94   &\textbf{0.938} (0.003)\\
& CT-Lasso adapt  &\textbf{0.063} (0.004) &10.0  &4.0  &1.00 &0.96   &\textbf{0.943} (0.003)\\
\hline
\end{tabular}
\end{small}
\end{table}
\example{3} (Grouped variables.) This example has $p=100$
predictors with
$\mathbf{\beta}^*=\{3,3,2.5,2.5,2,2,1.5,1.5,1,1,0,\ldots,0\}$. The
predictors are generated as $\mathbf{X}_j=Z_1 +
\sqrt{17/3}\epsilon_{x,j}$ for $j \in \{1,\ldots,10\}$,
$\mathbf{X}_j=Z_2 + \sqrt{1/19}\epsilon_{x,j}$ for $\mathbf{X}_j
\in \{11,\ldots,15\}$, and $\mathbf{X}_j=\epsilon_{x,j}$
otherwise, where $Z_1 \sim N(0,1)$, $Z_2 \sim N(0,1)$, and
$\epsilon_{x,j} \sim N(0,1)$ are independent. This creates
within-group correlations of $\mathbf{\Sigma}_{ij}=0.15$ for
$i,j\in \{1,\ldots,10\}$ and $\mathbf{\Sigma}_{ij}=0.95$ for
$i,j\in \{11,\ldots,15\}$.
$\sigma=15$ and SNR is approximately $1.1$.  This example presents
an interesting scenario where a group of significant variables are
mildly correlated and simultaneously a group of insignificant
variables are strongly correlated.

In Figure \ref{fig:best}(c), we see that the
covariance-thresholded lasso dominates generally in terms of variable selection. Similarly, Table
\ref{tab:ex3} shows that the covariance-thresholded lasso
does relatively well compared with other methods when
using cross-validation to select tuning parameters. Further, the
elastic net tends to have lower specificities than the
covariance-thresholded lasso methods. In the related scenario of
Example 4 in \citet*{Zou05}, where a group of significant
variables has strong within-group correlation and independent
otherwise, the performances of elastic net are similar to those of
covariance-thresholded lasso using soft thresholding, as both
methods regularize covariances with large magnitudes.

\begin{table}[!tb]\renewcommand{\baselinestretch}{1}
\caption{\label{tab:cv} Performances of cross-validation methods
based upon 200 replications of covariance-thresholded lasso with soft thresholding. $CV_{-}$ includes
additional variables up to 1 standard deviation of the minimum
cross-validation error; $CV_0$ selects $\lambda_n$ at the
minimum; and $CV_{+}$ discards
variables up to 1 standard deviation of the minimum. 
} \centering
\begin{small}
\begin{tabular}{|l|l||lll|lll|lll|}
\hline \emph{Example} & \emph{n} & \multicolumn{3}{c|}{\textit{$CV_{-}$}} & \multicolumn{3}{c|}{\textit{$CV_{0}$}}
& \multicolumn{3}{c|}{\textit{$CV_{+}$}} \\
& & \textit{sens} & \textit{spec} & \textit{$G$} &
\textit{sens} & \textit{spec} & \textit{$G$} & \textit{sens} &
\textit{spec} & \textit{$G$}\\
\hline Ex 1 & 20 & 0.60 & 0.78 & \textbf{0.667} (\textbf{0.007}) & 0.50 & 0.92 & 0.607 (0.009)
& 0.00 & 1.00 & 0.000 (0.086) \\

& 40 & 0.80 & 0.77 & \textbf{0.739} (\textbf{0.007}) & 0.60 & 0.92 &
0.699 (0.011) & 0.30 & 1.00 & 0.548 (0.019)\\

& 60 & 0.80 & 0.73 & 0.756 (\textbf{0.013}) & 0.70 & 0.93 &
\textbf{0.776} (\textbf{0.013}) & 0.40 & 1.00 & 0.632 (0.016)\\

& 80 & 0.90 & 0.73 & 0.789 (0.011) & 0.80 & 0.94 & \textbf{0.827}
(0.008) &
0.50 & 1.00 & 0.707 (\textbf{0.002})\\

\hline Ex 2& 20 & 0.33 & 0.71 & \textbf{0.465} (\textbf{0.008}) &
0.23 & 0.83 & 0.409 (0.012) & 0.15 & 0.89 & 0.362 (0.009)\\

& 40 & 0.50 & 0.56 & \textbf{0.485} (\textbf{0.006}) & 0.30 & 0.79
& 0.463 (0.009) &
0.20 & 0.86 & 0.414 (0.014)\\

& 60 & 0.55 & 0.54 & \textbf{0.504} (0.009) & 0.35 & 0.71 &
0.497 (\textbf{0.007}) &
0.30 & 0.79 & 0.473 (0.011) \\

& 80 & 0.65 & 0.48 & 0.505 (0.009) & 0.45 & 0.66 &
\textbf{0.521} (\textbf{0.007}) &
0.35 & 0.74 & 0.498 (0.008)\\

 \hline

Ex 3 & 20 & 0.70 & 0.77 & \textbf{0.700} (\textbf{0.013}) & 0.60 &
0.90 & 0.671 (0.015) &
0.20 & 0.99 & 0.433 (0.117)\\

& 40 & 0.90 & 0.76 & 0.815 (\textbf{0.007}) & 0.80 & 0.91 & \textbf{0.846}
(0.011) &
0.60 & 0.99 & 0.762 (0.025)\\

& 60 & 1.00 & 0.79 & 0.865 (0.006) & 0.90 & 0.93 & \textbf{0.922}
(\textbf{0.004}) &
0.80 & 0.99 & 0.872 (0.018) \\

& 80 & 1.00 & 0.80 & 0.882 (0.007) & 1.00 & 0.94 & \textbf{0.938} (0.003) &
0.80 & 1.00 & 0.894 (\textbf{0.002}) \\

\hline
\end{tabular}
\end{small}
\end{table}
\textbf{Methods of Cross-Validation. } We examine the modified
cross-validation presented in the beginning of this section.  In
Table \ref{tab:cv}, we summarize results of 3 variants of
cross-validation from covariance-thresholded lasso with soft thresholding.  Cross-validation by including additional variables up to 1 standard deviation of the minimum ($CV_{-}$), cross-validation by minimum validation error ($CV_{0}$), and cross-validation by discarding variables up to 1 standard deviation of the minimum ($CV_{+}$) are presented. The largest $G$ value and smallest bootstrapped standard deviations of $G$ among cross-validation methods are highlighted in boldface.

The results demonstrate the overwhelming pattern that the
proportion of relevant variables selected, or sensitivity,
decreases with $n$ under cross-validation. We
note that $CV_{+}$, as recommended in
\citet*{Hastie01}, does not work well in general for $n < p$. For
$n$ very small, $CV_{0}$ often selects too few
variables, whereas, for $n$ relatively large, $CV_{-}$ usually includes
too many irrelevant variables. Moreover,
when $n$ is very small, bootstrapped standard deviations of $G$
are usually the smallest for $CV_{-}$,
whereas, when $n$ is relatively large, $CV_{0}$ usually yields better standard deviations of $G$. These observations suggest the modified cross-validation that employs $CV_{0}$ when $n/\sqrt{p} > 5$ and
$CV_{-}$ when $n/\sqrt{p} < 5$.
\par \bigskip

\setcounter{chapter}{5}
\noindent {\bf 5. Real Data}
\par \smallskip

In this section, we compare the performance of
covariance-thresholded lasso with those of lasso, adaptive lasso, UST, scout(1,1), scout(2,1), and elastic net.  We apply the methods to 3 well-known data sets. For each data set, we randomly partition the data into a training
and a testing set. Tuning parameters are estimated using fivefold
cross-validations on the training set, and performances are
measured with the testing set. When $n/\sqrt{p}<5$, the modified
cross-validation described in Section 4 is used, where additional
variables are included up to 1 standard deviation of the
validation error at the minimum.
In order to avoid inconsistency of results due to randomization
\citep*{Bovelstad07}, we repeat the comparisons 100 times, each
with a different random partition of the training and testing set.
In Table \ref{tab:realdata}, we report median test MSE or classification error and number of
variables selected.  The smallest 3 test MSEs or classification errors are highlighted in boldface.  In addition, standard errors based on
500 bootstrapped re-samplings are reported in parentheses.

\textbf{Highway data. } Consider the highway accident data from an
unpublished master's paper by C. Hoffstedt and examined in
\citet*{Weisberg80}. The data set contains 39 observations, which
we divide randomly into $n=28$ and $nTest=11$ observations for the
training and testing set, respectively. The response is
$y$=accident rate per million vehicle miles. There are originally
9 predictors, and we further include quadratic and interaction
terms to obtain a total of $p=54$ predictors. The original
predictors are $X_1$=length of highway segment, $X_2$=average
daily traffic count, $X_3$=truck volume as a percentage of the
total volume, $X_4$=speed limit, $X_5$=width of outer shoulder,
$X_6$=number of freeway-type interchanges per mile, $X_7$=number
of signalized interchanges per mile, $X_8$=number of access points
per mile, and $X_9$=total number of lanes of traffic in both
directions.

\begin{table}[!tb]\renewcommand{\baselinestretch}{1}
\caption{\label{tab:realdata} Highway (\emph{$n$=28, $nTest$=11,
$p$=54}), CDI (\emph{$n$=308, $nTest$=132, $p$=90}), and Golub
microarray (\emph{$n$=38, $nTest$=34, $p$=1,000}) data performance
results based on 100 random partitions of training and testing
sets.
}
\centering
\begin{small}
\begin{tabular}{|l||lr|lr|lr|lr|}
\hline
\emph{Method} & \multicolumn{2}{c|}{\emph{Highway}} & \multicolumn{2}{c|}{\emph{CDI}} & \multicolumn{2}{c|}{\emph{Golub Microarray}}\\
& \emph{tMSE} & \emph{no.} & \emph{tMSE/$10^{10}$} & \emph{no.} & \emph{test error} & \emph{no.} \\
 \hline
Lasso             & 6.836 (0.917) & 24 (0.5) & 0.925 (0.225)    & 82 (1.8)     & 3.0 (0.383) & 37 (0.0)\\
Adapt Lasso       & 6.246 (0.577) & 22 (0.4) & 0.701 (0.263)    & 67.5 (4.0) & 3.0 (0.401) & 37 (0.0) \\
UST               & 12.948 (1.138) & 24 (0.9) & 1.562 (0.115)    & 20 (0.3) & \textbf{2.0} (0.447) & 198 (0.0)\\
Scout(1,1)        & \textbf{3.121} (0.172) & 20.5 (2.0) & NA & NA  & NA  & NA\\
Scout(2,1)        & \textbf{2.372} (0.292) & 17.5 (1.9) & \textbf{0.201} (0.014) & 6 (1.1)  & \textbf{1.0} (0.472)  & 194 (2.2)\\
Elastic net       & 6.165 (0.549) & 31 (2.2) & 0.216 (0.013)    & 22.5 (1.1) & 3.0 (0.336) & 26.5 (5.9)\\
CT-Lasso hard     & 5.400 (0.481) & 25 (1.4) & 0.226 (0.021)    & 35.5  (4.7) & 3.0 (0.388) & 21 (3.7)\\
CT-Lasso soft     & 3.480 (0.268) & 21 (2.5) & \textbf{0.185} (0.010)    & 21  (2.7) & 3.0 (0.388) & 24.5 (3.4)\\
CT-Lasso adapt    & \textbf{3.170} (0.486) & 19.5 (1.3) & \textbf{0.209} (0.015)    & 26 (2.4)    & \textbf{2.5} (0.476) & 36 (0.0)\\
\hline
\end{tabular}
\end{small}
\end{table}

Table \ref{tab:realdata} summarizes the results obtained.
Covariance-thresholded lasso methods with hard, soft, and adaptive
thresholding outperform the elastic net with 12\%, 44\%, and 49\%
reductions in median tMSE, respectively, and the lasso with 21\%,
49\%, and 54\% reductions in median tMSE, respectively.
The scout has the smallest tMSE.  We note that this may be due to scout's additional re-scaling step, in which it multiplies its initial estimates by a scalar $\hat{c} = \arg\min_c \|\mathbf{y}-c\mathbf{X}\hat{\mathbf{\beta}}\|^2$, as
explained in Section 4.

\textbf{CDI data.} Next, we consider the county demographic
information (CDI) data from the Geospatial and Statistical Data
Center of the University of Virginia and examined in
\citet*{Kutner05}. The data set contains 440 observations, which
we divide randomly into $n=308$ and $nTest=132$ observations for
the training and testing set, respectively. The response is
$y=$total number of crimes. There are originally 12 predictors,
and we further include quadratic and interaction terms to obtain a
total of $p=90$ predictors. The original predictors are $X_1$=land
area, $X_2$=population, $X_3$=percent 18-24 years old,
$X_4$=percent 65 years old or older, $X_5$=number of active
nonfederal physicians, $X_6$=number of hospital beds,
$X_7$=percent of adults graduated from high school, $X_8$=percent
of adults with bachelor's degree, $X_9$=percent below poverty
level income, $X_{10}$=percent of labor force unemployed,
$X_{11}$=per capita income, and $X_{12}$=total personal income.

Table \ref{tab:realdata} shows that the scout, elastic net, and
covariance-thresholded lasso dominate the lasso and adaptive lasso
in terms of prediction accuracy.  Covariance-thresholded lasso with soft thresholding performs the best with
80\% reduction in median tMSE from that
of the lasso. Adaptive lasso methods with relatively large
bootstrapped standard errors perform comparably to the lasso.

\textbf{Microarray data.} Finally, we consider the microarray data
from \citet*{Golub99}. This example seeks to distinguish acute
leukemias arising from lymphoid precursors (
ALL) and myeloid precursors (
AML). 
The data set contains 72 observations, which we divide randomly
into $n=38$ and $nTest=34$ observations for the training and
testing set, respectively. For the response $y$, we assign values
of 1 and -1 to ALL and AML, respectively. A classification rule is
applied for the fitted response such that ALL is represented if $y
\ge 0$ and AML otherwise. There are originally 7,129 predictors
from Affymetrix arrays. 
We use sure independence screening (SIS) with componentwise
regression, as recommended in \citet*{Fan08}, to first select
$p=1,000$ candidate genes. An early stop strategy is applied for
all methods at the 200th step, and cross-validation is performed
using the number of steps.

Table \ref{tab:realdata} presents results in terms of test errors
or the numbers of misclassifications out of 36 test samples. We
note that performances of the covariance-thresholded lasso methods
are comparable with those from the lasso, adaptive lasso, and
elastic net in terms of prediction accuracy. However,
covariance-thresholded lasso methods with hard and soft
thresholding select comparably less variables than the lasso,
adaptive lasso, and elastic net, whereas the scout severely over-selects with the number of variables selected close to the maximum of 200 due to early stopping. In the presence of comparable
prediction accuracy, this may suggest that covariance-thresholded
lasso can more readily differentiate between true and irrelevant
variables under high-dimensionality.

\par \bigskip

\setcounter{chapter}{6}
\noindent {\bf 6. Conclusion and Further Discussions}
\par \smallskip
In this paper, we have proposed the covariance-thresholded lasso,
a new regression method that stabilizes and improves the lasso for
variable selection by utilizing covariance sparsity, which is an
ubiquitous property in high-dimensional applications. The method
presents as an important marriage between methods of covariance
regularization \citep*{Bickel08} and variable selection. We have
shown theoretical studies and presented simulation and real-data
examples to indicate that our method can be useful in improving
variable selection performances, especially when $p \gg n$.

Furthermore, we note that there are many other variable selection procedures, such as the relaxed lasso \citep{Meinshausen07}, VISA \citep{Radchenko08}, etc., that may well be considered for comparison in Section 4 for the $n<p$ scenario.  However, due to limit in space, we restrict ourselves to only closely related methods in this paper.  We believe it can be interesting to further explore other methods for the $n<p$ scenario using modified cross-validation and best-possible selection of tuning parameters, and we hope to include them in future works.

Finally, sparse covariance-thresholding is a general procedure for
variable selection in high-dimensional applications. In this
paper, we applied covariance-thresholding specifically to the
lasso. Nonetheless, a myriad of variable selection methods, such
as the Dantzig
selector \citep*{Candes07}, 
SIS \citep*{Fan08}, etc., can also benefit by utilizing
covariance-thresholding to improve variable selection.  We believe
that results established in this paper will also be useful in
applying sparse covariance-thresholding for variable selection
methods other than the lasso.

\par \bigskip

\setcounter{chapter}{7}

\setcounter{lemma}{0}
\noindent {\bf 7. Appendix}
\par \smallskip
In this appendix, we first state and prove some preliminary lemmas
that will be used in later proofs. Lemma
\ref{Lemma:Sigma.lambda.ineq} gives the upper bounds of
$\mathbf{\hat \Sigma}^\nu_{CS}$ and $\mathbf{\hat
\Sigma}^\nu_{SS}$ as estimates of $\mathbf{\Sigma}_{CS}$ and
$\mathbf{\Sigma}_{SS}$, respectively. Lemma \ref{Lemma:Sigma.ineq}
gives the upper bound of any sample covariance matrix as an
estimate of its population counterpart. The rest of the appendix
is dedicated to the proofs of results in Section 3.1. The proofs of
results in Section 3.2, which we omit, are similar to those in
Section 3.1, except that $\nu$ is set to be 0 and Lemma
\ref{Lemma:Sigma.ineq} is used in place of Lemma
\ref{Lemma:Sigma.lambda.ineq}.

\par \bigskip

\noindent {\bf 7.1. Preliminary Lemmas}
\par \smallskip
\begin{lemma} \label{lemma:exp.ineq.general}
Suppose $(X_{k1}, X_{k2}, \ldots, X_{kp})$, $1\le k \le n$, are
independent and identically distributed random vectors with
$E(X_{kj}) = 0$, $E(X_{ki}X_{kj})=\sigma_{ij}$, and $ E
X_{kj}^{2d} \le d! M^d$ for $d\in \mathbb{N}\cup \{0\}$,
$M> 0$ and $1\le i, j \le p$. Let $\hat{\sigma}_{ij}={1\over
n}\sum_{k=1}^n X_{ki}X_{kj}$. Then, for $t_n = o(1)$,
\begin{equation}
P \left( |\hat \sigma_{ij} - \sigma_{ij}| > t_n \right) \le \exp(-
c n t_n^2),
\end{equation}
where $c$ is some constant depending only on $M$.
\end{lemma}

\noindent {\em Proof of Lemma \ref{lemma:exp.ineq.general}}

Let $Z_k = X_{ki} X_{kj} - \sigma_{ij}$. We apply the Bernstein's
Inequality (moment version) (see for example \citet{van96}) on the
series $\sum_{k = 1}^n Z_k$.

For $m \ge 1$, we have $ E|Z_k|^m = E |X_{ki} X_{kj} -
\sigma_{ij}|^m  \le  \sum_{d=0}^m \left({m \atop d}  \right)
|\sigma_{ij}|^{m-d} E |X_{ki} X_{kj}|^d$. By the moment conditions
in Lemma \ref{lemma:exp.ineq.general}, we have $|\sigma_{ij}| \le
M$ and $ E|X_{ki} X_{kj}|^d \le {1\over 2} \left( E X_{ki}^{2d} +
E X_{kj}^{2d} \right) \le d! M^d$. Therefore, $ E|Z_k|^m \le m!
M^m \sum_{d=0}^m \left({m \atop d}  \right) = m! (2 M)^m$, and
result follows by applying the moment version of Bernstein's
Inequality.  \qed

\begin{lemma} \label{Lemma:Sigma.lambda.ineq}
If $\nu$ is chosen to be greater than $C \sqrt{\log (s(p-s))} /
\sqrt{n}$ for some $C$ large enough, then
\begin{equation} \label{15.4}
\left\|\hat{\mathbf{\Sigma}}^\nu_{CS}-\mathbf{\Sigma}_{CS}\right\|_\infty
\le O_p \left( \nu d^*_{CS} \right) +  O_p \left( d^*_{C S}
\sqrt{\log(s(p-s))} / \sqrt{n} \right).
\end{equation}
If $\nu$ is chosen to be greater than $C \sqrt{\log s} / \sqrt{n}$
for some $C$ large enough, then
\begin{equation} \label{15.5}
\left\|\hat{\mathbf{\Sigma}}^\nu_{SS}
-\mathbf{\Sigma}_{SS}\right\|_\infty \le O_p \left( \nu d^*_{SS}
\right) +  O_p \left(  d^*_{S S} \sqrt{2 \log s} / \sqrt{n}
\right).
\end{equation}

\end{lemma}

The proof is similar to that of Theorem 1 in \citet{Bickel08} and
Theorem 1 in \citet{Rothman08}, and, thus, it is omitted to save
space. The detailed proof can be found in the supplementary document.

\begin{lemma} \label{Lemma:Sigma.ineq}
Let $A$ and $B$ be two arbitrary subsets of $\{1,2,\ldots, p\}$,
and let ${\Sigma}_{AB}=({\sigma}_{ij})_{i\in A, j\in B}$ and
$\hat{\Sigma}_{AB}=(\hat{\sigma}_{ij})_{i\in A, j\in B}$. Further,
let $a$ be the cardinality of $A$ and $b$ the cardinality of $B$.
Suppose $a$ and $b$ satisfy $\sqrt{\log (ab)} / \sqrt{n} \to 0$ as
$n\to \infty$. Then
\begin{equation*}
\| \mathbf{\hat \Sigma}_{AB} - \mathbf{\Sigma}_{AB} \|_\infty =
O_p \left( b \sqrt{\log (ab)} / \sqrt{n}  \right).
\end{equation*}
\end{lemma}

\noindent
{\em Proof of Lemma \ref{Lemma:Sigma.ineq}}

Since
\begin{eqnarray*}
P\left(\|\hat{\mathbf{\Sigma}}_{AB} -
\mathbf{\Sigma}_{AB}\|_{\infty}
> t \right) \le
\sum_{i \in A} \sum_{j \in B} P\left( \left|\hat{\sigma}_{ij} -
\sigma_{ij} \right| > t/b \right) \le  a \cdot b \cdot \exp(- c n
t^2/ b^2),
\end{eqnarray*}
for $t/b = o(1)$ by Lemma \ref{lemma:exp.ineq.general}, the
result follows. \qed
\par \bigskip

\noindent {\bf 7.2. Proof of Lemma \ref{Lemma:KKT.et}}
\par \smallskip
By the KKT conditions, the solution of (\ref{def:CT-Lasso})
satisfies $ \mathbf{\hat \Sigma}_\nu \hat \beta^\nu - {1\over n}
\mathbf{X}^T y + \lambda_n \hat z = 0$, where $\hat z$ is the
sub-gradient of $\|\hat \beta^\nu \|_1$, that is, $\hat z=
\partial \|\hat \beta^\nu \|_1$. Plugging in $y =\mathbf{X}
\beta^* + \epsilon$, we have
\begin{equation} \label{4}
\mathbf{\hat \Sigma}_\nu (\hat \beta^\nu - \beta^* ) +
(\mathbf{\hat \Sigma}_\nu - \mathbf{\hat \Sigma}) \beta^* -
{1\over n} \mathbf{X}^T \epsilon + \lambda_n \hat z = 0.
\end{equation}
It is easy to see that $sgn(\hat \beta^\nu) = sgn(\beta^*)$ holds
if $\hat \beta_S^\nu \ne 0$, $\hat \beta_{C}^\nu = 0$, $\hat z_S =
sgn(\beta^*_S)$, and $|\hat z_{C}| \le 1$. Therefore, based on
(\ref{4}), the conditions for $sgn(\hat \beta^\nu ) = sgn(\hat
\beta^*)$ to hold are
\begin{equation} \label{4.1}
\mathbf{\hat \Sigma}^\nu_{SS} (\hat \beta^\nu_S - \beta^*_S ) +
(\mathbf{\hat \Sigma}^\nu_{SS} - \mathbf{\hat \Sigma}_{SS})
\beta^*_S - {1\over n} \mathbf{X}^T_S \epsilon = - \lambda_n
sgn(\beta^*_S),
\end{equation}
\begin{equation} \label{4.4}
sgn(\hat \beta_S^\nu) = sgn(\beta^*_S),
\end{equation}
\begin{equation} \label{4.2}
\left\| \mathbf{\hat \Sigma}^\nu_{CS} (\hat \beta^\nu_S -
\beta^*_S ) + (\mathbf{\hat \Sigma}^\nu_{CS} - \mathbf{\hat
\Sigma}_{CS}) \beta^*_S - {1\over n} \mathbf{X}^T_{C} \epsilon
\right\|_\infty \le \lambda_n.
\end{equation}
Solving (\ref{4.1}) for $\hat \beta_S^\nu$ under the assumption
$\Lambda_{min}\left(\mathbf{\hat \Sigma}_{SS}^\nu \right)
> 0$, we have
\begin{equation} \label{4.3}
\hat \beta_S^\nu = \beta_S^* + (\mathbf{\hat
\Sigma}_{SS}^\nu)^{-1} \left({1\over n} \mathbf{X}_S^T \epsilon -
\lambda_n sgn(\beta^*_S) - \left( \mathbf{\hat \Sigma}_{SS}^\nu -
\mathbf{\hat \Sigma}_{SS} \right) \beta^* \right).
\end{equation}
Substituting (\ref{4.3}) into the left-hand side of (\ref{4.2})
and further decomposing the resulting equation, we have
\begin{eqnarray*}
& & \left\|\mathbf{\hat \Sigma}_{CS}^\nu (\mathbf{\hat
\Sigma}_{SS}^\nu)^{-1} \left({1\over n} \mathbf{X}_S^T \epsilon -
\lambda_n sgn(\beta^*_S) - \left(\mathbf{\hat \Sigma}_{SS}^\nu -
\mathbf{\hat \Sigma}_{SS} \right)\beta^*_S \right) + \left(
\mathbf{\hat \Sigma}_{CS}^\nu - \mathbf{\hat \Sigma}_{CS}
\right)\beta^*_S - {1\over n} \mathbf{X}_{C}^T
\epsilon \right\|_\infty \\
& \le & \left\|\mathbf{\hat \Sigma}_{CS} (\mathbf{\hat
\Sigma}_{SS})^{-1} \right\|_\infty  \left( \left\| {1\over n}
\mathbf{X}_S^T \epsilon\right\|_\infty + \lambda_n +
\left\|\left(\mathbf{\hat \Sigma}_{SS}^\nu - \mathbf{\hat
\Sigma}_{SS} \right)\beta^*_S
\right\|_\infty \right) \\
& + & \left\|\left( \mathbf{\hat \Sigma}_{CS}^\nu - \mathbf{\hat
\Sigma}_{CS} \right)\beta^*_S \right\|_\infty + \left\|{1\over n}
\mathbf{X}_{C}^T
\epsilon \right\|_\infty \\
& \le & \left\|\mathbf{\hat \Sigma}_{CS} (\mathbf{\hat
\Sigma}_{SS})^{-1} \right\|_\infty  \left( \left\| {1\over n}
\mathbf{X}_S^T \epsilon\right\|_\infty + s \nu \overline{\rho} +
\lambda_n \right) +  s \nu \overline{\rho} + \left\|{1\over n}
\mathbf{X}_{C}^T \epsilon \right\|_\infty,
\end{eqnarray*}
where the last inequality is obtained by
\begin{equation} \label{2}
\left\| \left(\mathbf{\hat \Sigma}_{SS}^\nu - \mathbf{\hat
\Sigma}_{SS} \right) \beta^*_S \right\|_\infty \le \left\|
\mathbf{\hat \Sigma}_{SS}^\nu - \mathbf{\hat \Sigma}_{SS}
\right\|_\infty \|\beta^*_S\|_\infty \le s \nu \overline{\rho},
\end{equation}
\begin{equation*} \label{2.1}
\left\| \left(\mathbf{\hat \Sigma}_{CS}^\nu - \mathbf{\hat
\Sigma}_{CS} \right) \beta^*_S \right\|_\infty \le \left\|
\mathbf{\hat \Sigma}_{CS}^\nu - \mathbf{\hat \Sigma}_{CS}
\right\|_\infty \left\| \beta^*_S \right\|_\infty \le s \nu
\overline{\rho}.
\end{equation*}
Then, condition (\ref{Ineq:et.KKT.1}) is sufficient for
(\ref{4.2}) to hold.

Next, we derive (\ref{Ineq:et.KKT.2}). By (\ref{4.3}), (\ref{4.4})
is implied by
\begin{equation} \label{5}
\left\| (\mathbf{\hat \Sigma}_{SS}^\nu)^{-1} \right\|_\infty
\left( \left\| \left( \mathbf{\hat \Sigma}_{SS}^\nu - \mathbf{\hat
\Sigma}_{SS} \right) \beta^* \right\|_\infty + \left\|{1\over n}
\mathbf{X}_S^T \epsilon \right\|_\infty + \lambda_n \right) <
\underline{\rho}.
\end{equation}
Plugging in the upper bound of $\| (\mathbf{\hat \Sigma}_{SS}^\nu
- \mathbf{\hat \Sigma}_{SS} ) \beta^*_S \|_\infty $ in (\ref{2}),
it is straightforward to see that (\ref{Ineq:et.KKT.2}) is
sufficient for (\ref{4.4}) to hold. \qed
\par \bigskip

\noindent {\bf 7.3. Proof of Lemma \ref{Lemma:nonsingular.et}}
\par \smallskip
For any $v$ with $\| v \| = 1$,
\[
v^T \hat{\mathbf{\Sigma}}_{SS}^\nu v \ge
\Lambda_{min}\left(\mathbf{\Sigma}_{SS} \right) - \|
\hat{\mathbf{\Sigma}}_{SS}^\nu - \mathbf{\Sigma}_{SS}\| \ge
\Lambda_{min}\left(\mathbf{\Sigma}_{SS} \right) - \|
\hat{\mathbf{\Sigma}}_{SS}^\nu - \mathbf{\Sigma}_{SS}\|_{\infty},
\]
and, when choosing $\nu = C \sqrt{\log s} / \sqrt{n}$ for some
$C>0$,
\begin{equation} \label{7}
\| \hat{\mathbf{\Sigma}}_{SS}^\nu -
\mathbf{\Sigma}_{SS}\|_{\infty} \le O_p\left( d^*_{SS} \sqrt{\log
s} / \sqrt{n} \right)
\end{equation}
by Lemma \ref{Lemma:Sigma.lambda.ineq}. Therefore, the result
follows under the condition (\ref{cond:d*.1}). \qed
\par \bigskip

\noindent {\bf 7.4. Proof of Lemma \ref{Lemma:Irrep.et}}
\par \smallskip
To derive the upper bound of $\| (\hat{\mathbf{\Sigma}}_{SS}^\nu
)^{-1} \|_{\infty}$, we perform the following decomposition,
\begin{eqnarray} \label{7.5}
\left\| \left(\hat{\mathbf{\Sigma}}_{SS}^\nu \right)^{-1}
\right\|_{\infty} & \le & \left\| \left(\mathbf{\Sigma}_{SS}
\right)^{-1} \right\|_{\infty} + \left\|
\left(\hat{\mathbf{\Sigma}}_{SS}^\nu \right)^{-1} -
\left(\mathbf{\Sigma}_{SS} \right)^{-1} \right\|_\infty.
\end{eqnarray}
Because
\begin{eqnarray*}
\left\| \left(\hat{\mathbf{\Sigma}}_{SS}^\nu \right)^{-1} -
\left(\mathbf{\Sigma}_{SS}\right)^{-1} \right\|_\infty & \le &
\left\| \left(\mathbf{\Sigma}_{SS}\right)^{-1} \right\|_\infty
\left\| \left(\hat{\mathbf{\Sigma}}_{SS}^\nu \right)^{-1}
\right\|_\infty \left\| \hat{\mathbf{\Sigma}}_{SS}^\nu -
\mathbf{\Sigma}_{SS} \right\|_\infty
 \nonumber \\
& \le & \bar{D} \left( \left\|
\left(\mathbf{\Sigma}_{SS}\right)^{-1} \right\|_\infty +  \left\|
\left(\hat{\mathbf{\Sigma}}_{SS}^\nu \right)^{-1} -
\left(\mathbf{\Sigma}_{SS}\right)^{-1} \right\|_\infty \right)
\left\| \hat{\mathbf{\Sigma}}_{SS}^\nu - \mathbf{\Sigma}_{SS}
\right\|_\infty \\
& = & \bar{D}^2 \left\| \hat{\mathbf{\Sigma}}_{SS}^\nu -
\mathbf{\Sigma}_{SS} \right\|_\infty  + \bar{D} \left\|
\left(\hat{\mathbf{\Sigma}}_{SS}^\nu \right)^{-1} -
\left(\mathbf{\Sigma}_{SS}\right)^{-1} \right\|_\infty \left\|
\hat{\mathbf{\Sigma}}_{SS}^\nu - \mathbf{\Sigma}_{SS}
\right\|_\infty,
\end{eqnarray*}
where the second inequality is obtained by (\ref{7.5}), we have
\begin{equation} \label{7.2}
\left\| \left(\hat{\mathbf{\Sigma}}_{SS}^\nu \right)^{-1} -
\left(\mathbf{\Sigma}_{SS}\right)^{-1} \right\|_\infty  \le
{\bar{D}^2 \left\| \hat{\mathbf{\Sigma}}_{SS}^\nu -
\mathbf{\Sigma}_{SS} \right\|_\infty \over 1 - \bar{D} \left\|
\hat{\mathbf{\Sigma}}_{SS}^\nu - \mathbf{\Sigma}_{SS}
\right\|_\infty} \le O_p \left({\bar{D}^2 d_{SS}^* \sqrt{\log
(p-s)} \over \sqrt{n}} \right)
\end{equation}
where the last inequality is derived by choosing $\nu =
C\sqrt{\log (p-s)} / \sqrt{n}$, applying (\ref{15.5}) in Lemma
\ref{Lemma:Sigma.lambda.ineq}, and using the condition
(\ref{cond:d*.2}). Combining (\ref{7.5}), (\ref{7.2}), and
condition (\ref{cond:d*.2}), we have
\begin{equation} \label{7.3}
\left\| \left(\hat{\mathbf{\Sigma}}_{SS}^\nu \right)^{-1}
\right\|_{\infty} \le O_p \left( \bar{D} \right).
\end{equation}

For (\ref{Irrep1}), we decompose $\hat{\mathbf{\Sigma}}_{CS}^\nu
(\hat{\mathbf{\Sigma}}_{SS}^\nu)^{-1}$ into three terms as
follows:
\begin{eqnarray*} \label{9.5}
\hat{\mathbf{\Sigma}}_{CS}^\nu \left(
\hat{\mathbf{\Sigma}}_{SS}^\nu \right)^{-1} & =
&\hat{\mathbf{\Sigma}}_{CS}^\nu
\left[\left(\hat{\mathbf{\Sigma}}_{SS}^\nu \right)^{-1}
-\left(\mathbf{\Sigma}_{SS} \right)^{-1}\right] +\left[
\hat{\mathbf{\Sigma}}_{CS}^\nu -\mathbf{\Sigma}_{CS} \right]
\left(\mathbf{\Sigma}_{SS}\right)^{-1}
+\mathbf{\Sigma}_{CS}\left(\mathbf{\Sigma}_{SS} \right)^{-1}\\
& = & \text{I}+\text{II}+\text{III}.
\end{eqnarray*}
By condition (\ref{Cond:Sigma2}),
$\left\|\text{III}\right\|_{\infty}\le 1-\epsilon$. Therefore, it
is enough to show that $\|\text{I}\|_\infty + \|\text{II}\|_\infty
\le \epsilon / 2$ with probability going to $1$.

For $\|\text{II}\|_{\infty}$, we have
\begin{eqnarray} \label{7.6}
\left\|\text{II}\right\|_{\infty} & \le &
\|(\mathbf{\Sigma}_{SS})^{-1}\|_{\infty}
\|\hat{\mathbf{\Sigma}}_{CS}^\nu - \mathbf{\Sigma}_{CS}\|_{\infty}
=  \bar{D} \cdot \|\hat{\mathbf{\Sigma}}_{CS}^\nu -
\mathbf{\Sigma}_{CS}\|_{\infty},
\end{eqnarray}
and, when choosing $\nu = C \sqrt{\log(s(p-s))} / \sqrt{n}$,
\begin{equation} \label{7.8}
\left\|\hat{\mathbf{\Sigma}}^\nu_{CS}-\mathbf{\Sigma}_{CS}\right\|_\infty
\le O_p \left(d^*_{C S} \sqrt{\log(s(p-s))} / \sqrt{n} \right)
\end{equation}
by Lemma \ref{Lemma:Sigma.lambda.ineq}. For
$\|\text{I}\|_{\infty}$, we have
\begin{eqnarray}
\|\text{I}\|_{\infty} & \le & \|\hat{\mathbf{\Sigma}}_{CS}^\nu
\|_\infty \left\| \left(\hat{\mathbf{\Sigma}}_{SS}^\nu
\right)^{-1} -\left(\mathbf{\Sigma}_{SS}
\right)^{-1} \right\|_\infty \nonumber \\
& \le & \left( \| \mathbf{\Sigma}_{CS}\|_\infty +
\|\hat{\mathbf{\Sigma}}_{CS}^\nu - \mathbf{\Sigma}_{CS}\|_{\infty}
\right) \left\| \left(\hat{\mathbf{\Sigma}}_{SS}^\nu
\right)^{-1} -\left(\mathbf{\Sigma}_{SS} \right)^{-1} \right\|_\infty  \label{7.7}\\
& \le & O_p \left( \bar{D}^2 d^*_{CS} d^*_{SS} \sqrt{\log (s
(p-s))} / \sqrt{n} \right) \nonumber
\end{eqnarray}
by (\ref{7.8}), (\ref{7.2}), and (\ref{Cond:(d*np)}).

In summary, we have $ P \left( \|\text{I}\|_\infty +
\|\text{II}\|_\infty \le \epsilon / 2 \right) \to 1 $ under the
condition (\ref{Cond:(d*np)}). This completes the proof. \qed
\par \bigskip

\noindent {\bf 7.5. Proof of Theorem {\ref{Thm:et.vsc}}}
\par \smallskip
First, we consider $\left\| {1\over n} \mathbf{X}_{C}^T
\epsilon\right\|_\infty$ and $\left\| {1\over n} \mathbf{X}_{S}^T
\epsilon\right\|_\infty$, which appear in (\ref{Ineq:et.KKT.1})
and (\ref{Ineq:et.KKT.2}), respectively. Since $\epsilon \sim N(0,
\sigma^2)$, then, when $\mathbf{X}$ is fixed, by standard results
on the extreme value of multivariate normal, we have
\begin{equation} \label{Eqn:mathbfX*error1}
\left\| {1\over n} \mathbf{X}_{C}^T \epsilon\right\|_\infty =
O_p\left( \sigma \sqrt{2 (\max_{j} \hat \sigma_{jj}) \log (p-s)} /
\sqrt{n} \right),
\end{equation}
\begin{equation} \label{Eqn:mathbfX*error2}
\left\| {1\over n} \mathbf{X}_S^T \epsilon\right\|_\infty = O_p
\left( \sqrt{2 (\max_{j} \hat \sigma_{jj}) \log s} / \sqrt{n}
\right).
\end{equation}
By Lemma \ref{lemma:exp.ineq.general},
\begin{eqnarray*}
P \left( \max_{j} \hat \sigma_{jj}  > M + t \right) \le
\sum_{j=1}^p P \left( \hat \sigma_{jj}  > M + t \right) \le
\sum_{j=1}^p P \left( \hat \sigma_{jj} - \sigma_{jj} > t \right)
\le p \cdot \exp(-nt^2 / 4)
\end{eqnarray*}
for $t = o(1)$, and, thus, $P \left(\max_{j} \hat \sigma_{jj} \le
M \right) \to 1$. Therefore,
\begin{equation} \label{ineq:Xepslion}
\left\| {1\over n} \mathbf{X}_{C}^T \epsilon\right\|_\infty =
O_p\left(  \sqrt{ \log (p-s)} / \sqrt{n} \right), \qquad \left\|
{1\over n} \mathbf{X}_S^T \epsilon\right\|_\infty = O_p \left(
\sqrt{\log s} / \sqrt{n} \right).
\end{equation}

Now, we sum up the results in Lemma \ref{Lemma:KKT.et},
\ref{Lemma:nonsingular.et}, \ref{Lemma:Irrep.et}, and
(\ref{ineq:Xepslion}). Under conditions  (\ref{Cond:X moments}),
(\ref{Cond:Sigma1}), (\ref{Cond:Sigma2}), (\ref{Cond:(d*np)}),
(\ref{Cond:dim.1}), and the choice of $\nu = C\sqrt{\log (p-s)} /
\sqrt{n}$ for some $C$ and $\lambda_n$ as in
(\ref{cond:lambda.et}), both (\ref{Ineq:et.KKT.1}) and
(\ref{Ineq:et.KKT.2}) hold with probability going to $1$. \qed
\par \bigskip

\noindent {\bf 7.6. Outline of the Proof of Theorem \ref{thm:piecewiseLinear}}
\par \smallskip
To circumvent the problem of having a non-differentiable penalty function, we reformulate the optimization problem in (\ref{def:CT-Lasso}) as the following,
\[
\begin{array}{c}
\arg\min_{\beta^+, \beta^-} \,
\frac{1}{2}(\beta^+ - \beta^-)^T \hat{\mathbf{\Sigma}}_\nu (\beta^+ - \beta^-) - (\beta^+ - \beta^-)^T \left({1\over
n}\mathbf{X}^T \mathbf{y} \right), \\
\text{s. t.} \quad \beta_j^- \ge 0 \, \forall j, \quad
\beta_j^+ \ge 0 \, \forall j, \quad
\sum_j (\beta_j^+ + \beta_j^-) \le t.
\end{array}
\]
Consider the Lagrangian primal function for the above formulation,
\[
\frac{1}{2}(\beta^+ - \beta^-)^T \hat{\mathbf{\Sigma}}_\nu (\beta^+ - \beta^-) - (\beta^+ - \beta^-)^T \left({1\over
n}\mathbf{X}^T \mathbf{y} \right) + \lambda \sum_{j=1}^p (\beta_j^+ + \beta_j^-) - \sum_{j=1}^p
\lambda_j^+ \beta_j^+ - \sum_{j=1}^p \lambda_j^- \beta_j^-.
\]
Let $\beta = \beta^+ - \beta^-$.  We obtain the following first-order conditions,
\[
\begin{array}{cc}
\frac{1}{n}\mathbf{x}_j^T \mathbf{y} - (\hat{\mathbf{\Sigma}}_\nu)_j^T \beta - \lambda + \lambda_j^+ = 0, &
\frac{1}{n}\mathbf{x}_j^T \mathbf{y} - (\mathbf{\Sigma}_\nu)_j^T \beta
+ \lambda - \lambda_j^- = 0, \\
\lambda_j^+\beta_j^+ = 0, & \lambda_j^-\beta_j^- = 0.
\end{array}
\]
These conditions can be verified, as in \citep*{Rosset07}, to imply the facts,
\[
|\frac{1}{n}\mathbf{x}_j^T \mathbf{y} - (\hat{\mathbf{\Sigma}}_\nu)_j^T \beta| < \lambda \, \Longrightarrow \, \beta_j =0 \quad \text{and} \quad
\beta_j \neq 0 \, \Longrightarrow \, |\frac{1}{n}\mathbf{x}_j^T \mathbf{y} - (\hat{\mathbf{\Sigma}}_\nu)_j^T \beta| = \lambda.
\]

When $\hat{\mathbf{\Sigma}}_\nu$ is semi-positive definite, first-order conditions are enough to provide a global solution, which is unique if all eigenvalues are positive.  However, when there exist eigenvalues of $\hat{\mathbf{\Sigma}}_\nu$ that are negative, a second-order condition, in addition to first-order ones, is required to guarantee that a point $\beta$ is a local minimum.  Assume strict complementarity $\beta_j=0 \Longrightarrow \lambda_j^+ > 0\text{ and }\lambda_j^- > 0$,
which holds with high probability as regression methods rarely yield zero-valued coefficient estimate without penalization.  We see that $\mathcal{K} = \{z \doteq z^+ - z^- \neq 0: z_j^+=0\text{ and }z_j^-=0\text{ for }\beta_j=0 \}$ covers the set of feasible directions in Theorem 6, \citet{McCormick76}.   Let $\mathcal{A} = \{j : \beta_j \neq 0\}$.  By Theorem 6, \citet{McCormick76}, a solution $\beta$ is a local minimum if for every $z \in \mathcal{K}$
\[
z^T (\hat{\mathbf{\Sigma}}_\nu) z = (z_{\mathcal{A}})^T (\hat{\mathbf{\Sigma}}_\nu)_{\mathcal{A}} \, z_{\mathcal{A}} > 0.
\]
Furthermore, we note that the solution $\beta$ is global if $|\mathbf{x}_j^T \mathbf{y}/n| < \lambda$ for all $j \notin \mathcal{A}$ in addition to $(\hat{\mathbf{\Sigma}}_\nu)_{\mathcal{A}}$ being positive definite.  This follows from facts implied by first-order conditions.

Algorithm for computing piecewise-linear solutions for the covariance-thresholded lasso is derived by further manipulating the first-order conditions as in the proof for Theorem 2 in \citet{Rosset07}.

\par \bigskip

\noindent {\large\bf Acknowledgment}
\par \smallskip
The authors are grateful to Jayanta K. Ghosh and Jian Zhang for helpful comments and discussions. Furthermore, we thank the associate editor and two referees who have been very generous in providing us with helpful suggestions. Z. John Daye is supported by
Purdue Research Foundation Fellowship. Computing resources and
support were provided by the Department of Statistics, Purdue
University, and the Rosen Center for Advanced Computing (RCAC) of
Information Technology at Purdue.
\par \bigskip

\bibliography{etbib}

\vskip .65cm \noindent Department of Statistics, Purdue
University, West Lafayette, IN 47906, U.S.A. \vskip 2pt \noindent
E-mail: xingejeng@gmail.com \vskip 2pt \noindent Department of
Statistics, Purdue University, West Lafayette, IN 47906, U.S.A.
\vskip 2pt \noindent E-mail: zhongyindaye@gmail.com \vskip 2pt
\end{document}